\documentclass[12pt,A4paper]{article}
\usepackage[explicit]{titlesec}
\usepackage{hyphenat}
\usepackage{ragged2e}
\usepackage{amssymb,amsmath}
\usepackage{bbm}
\usepackage{graphicx,psfrag,epsf}
\usepackage{enumerate}
\usepackage[square,sort,comma,numbers]{natbib}
\usepackage{url}
\usepackage{algorithm2e}
\usepackage{multirow}
\usepackage{color}
\usepackage{float}
\usepackage{tikz}
\usetikzlibrary{arrows,decorations.pathmorphing,backgrounds,fit,positioning,shapes.symbols,chains}
\usepackage{geometry}
\usepackage{soul}
\usepackage{xr} 
\setul{.6pt}{.4pt}

\newtheorem{prop}{Proposition}

\titleformat{\section}
  {\normalfont}{\thesection}{1em}{\MakeUppercase{\textbf{#1}}}
\titlespacing\section{0pt}{0pt}{-10pt}
\titleformat{\subsection}
  {\normalfont}{\thesubsection}{1em}{\textit{#1}}
\titlespacing\subsection{0pt}{0pt}{-8pt}

\makeatletter
\newcommand\sixteen{\@setfontsize\sixteen{16pt}{6}}
\renewcommand{\maketitle}{\bgroup\setlength{\parindent}{0pt}
\begin{flushleft}
\vspace{-.375in}
\sixteen\bfseries \@title
\medskip
\end{flushleft}
\textit{\@author}
\egroup}
\makeatother


\title{Probabilistic reconstruction of truncated particle trajectories on a closed surface}
\author{Yunjiao Lu$^{1,2,\$}$, Pierre Hodara$^{1,\$}$, Charles Kervrann$^{2}$, Alain Trubuil$^{1}$* \vspace{0.5cm} \\
$^{1}$INRA, UR 1404, MaIAGE, Universit\'e Paris-Saclay, Jouy-en-Josas, France\\
$^{2}$Inria, Centre de Recherche Bretagne-Atlantique, EPC SERPICO, Rennes, France \vspace{0.5cm} \\
\textsuperscript{\$}these authors contributed equally to this work\\
\textsuperscript{*}corresponding author
}

\begin{document}
\justify
\vspace*{.01 in}
\maketitle
\vspace{.12 in}

\section*{abstract}
\vspace{10pt}
Investigation of dynamic processes in cell biology very often relies on the observation in two dimensions of 3D biological  processes. Consequently, the data are partial and statistical methods and models are required to recover the parameters describing the dynamical processes. In the case of molecules moving over the 3D surface, such as proteins on walls of bacteria cell, a large portion of the 3D surface is not observed in 2D-time microscopy. It follows that biomolecules may disappear for a period of time in a region of interest, and then reappear later. Assuming Brownian motion with drift, we address the mathematical problem of the reconstruction of biomolecules trajectories on a cylindrical surface. A subregion of the cylinder is typically recorded during the observation period, and biomolecules may appear or disappear in any place of the 3D surface. The performance of the method is demonstrated on simulated particle trajectories that mimic MreB protein dynamics observed in 2D time-lapse fluorescence microscopy in rod-shaped bacteria.

\section{Introduction}
\begin{figure}[t]
    \centering
    \includegraphics[scale = 0.185]{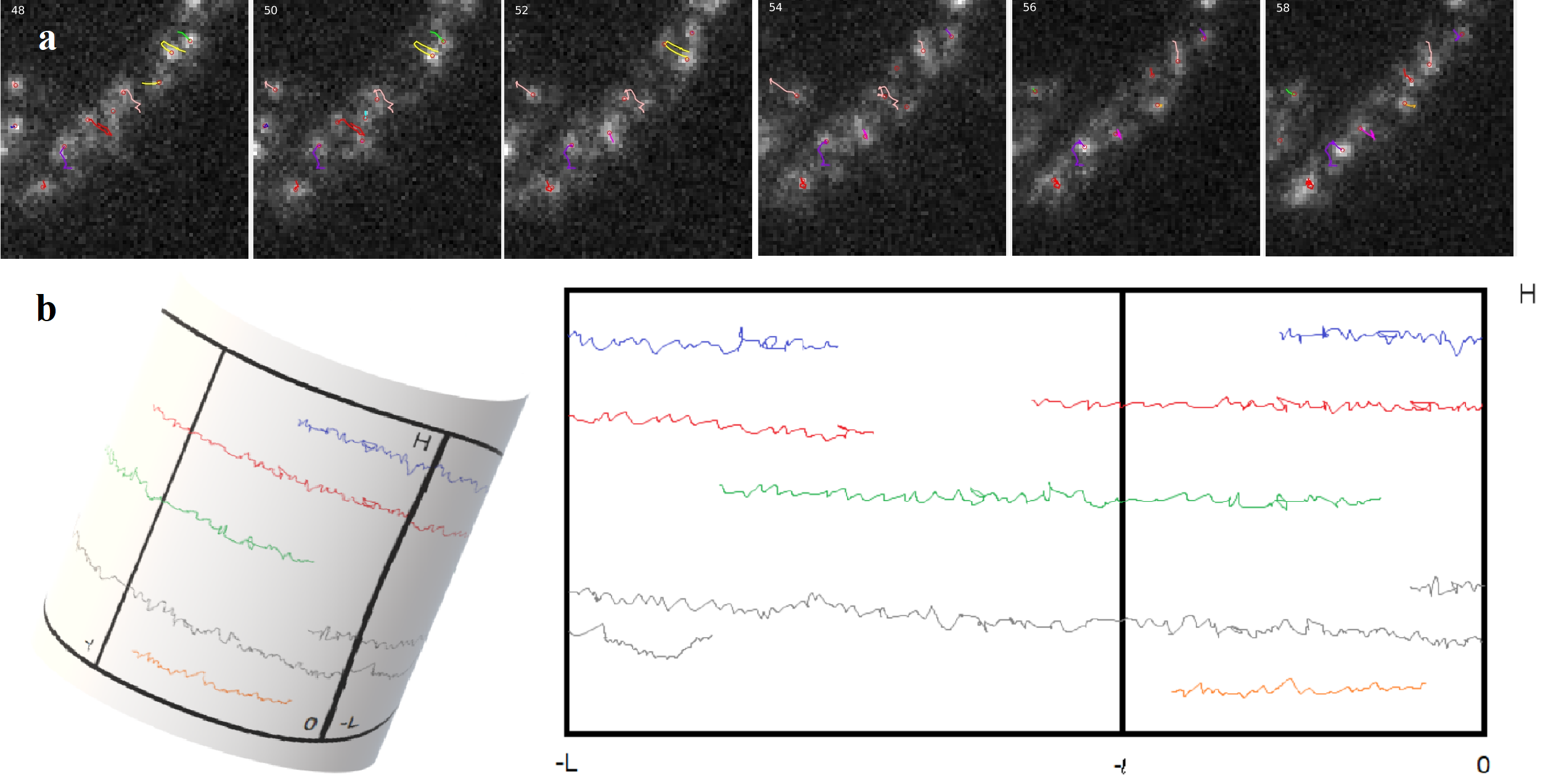}
    \caption{(a): Several consecutive images from a real TIRFM movie\cite{axelrod1984total}. Tracks are superposed on the images.(b) left:  Illustration of trajectories observed during recorded time $[0, T_S]$ on the surface of a cylinder. Only the motions inside the ROO $]-l,0[  \times [0,H]$ can be observed, even though the dynamics happen on the whole surface; right: Representation of the dynamics on a 2D unwrapped surface $]-L,0[  \times [0,H].$ The objective is to recover the dynamics on the whole surface from the partial observations, by coordinating the inputs through $\{-l\}\times [0,H]$ and the outputs through $\{0\}\times [0,H]$ in a movie during $T_S$, taking into account particles birth and death events.}    
    \label{fig:illus_partialobservation}
\end{figure}

In 2D and 3D live-cell imaging, spatiotemporal events and biomolecule dynamics are frequently observed with an incomplete field of view. Very often these observations are related to regions of observation (ROO) inside a tissue, a cell, or in the neighborhood of membranes. Nevertheless, it is quite unusual to analyze 3D dynamics of biomolecules or events occurring on a closed surface and observed on a 2D plane. Our work is motivated by the study of dynamics of  MreB proteins, moving close to the inner membrane during cell wall construction in rod-shaped bacteria (\cite{billaudeau2017contrasting}, \cite{van2018recent}). Its dynamics can only be observed in a small region and are recorded as 2D time-lapse movies (Fig. \ref{fig:illus_partialobservation}a). As for 3D image acquisition, even it can solve the problem of partial observation, is not always appropriate, especially if the objective is to capture fast and temporally short events as described in \cite{billaudeau2017contrasting}. The frame rate adapted to the scale of dynamics may be too high when compared to the period of time to acquire temporal series of 3D volume (\cite{boulanger2014fast} and \cite{cornilleau2020methods}).

To our knowledge, identifying re-entrance events of the same entities inside the ROO is not addressed in the literature. In experimental data, when the unobserved region represents a significant part of the entire surface, a complete description of the dynamics on these closed surfaces becomes of paramount importance for deciphering the mechanisms of some processes. In our study of the regulation of the dynamics of MreB protein, as inputs, we consider a set of trajectories estimated by tracking  algorithms (e.g. \cite{jaqaman2008robust}, \cite{chenouard2013multiple}). These tracking algorithms are very sophisticated and allow us to handle large sets of particles, different stochastic dynamical models \cite{blom1988interacting}, \cite{bressloff2013stochastic}, and observation models \cite{genovesio2006multiple}, \cite{li2001target}. They take into account birth/death events, and/or split/merge events. Particles may be unobserved or undetected for short periods of time, especially in 2D+time microscopy. However, none computational or statistical method manages the situation corresponding to a large hidden region inside the region of interest. Also, the identification of particles leaving the ROO through one border of the domain and re-entering from a far border is not addressed. Our objective is then to provide a generic approach to tackle the problem of the reconstruction of particle trajectories observed on a small part of a closed surface as illustrated in Fig. \ref{fig:illus_partialobservation}b. 

In this paper, we focus on the design and evaluation of a self-contained mathematical framework to tackle the reconstruction of particle trajectories on cylindrical surfaces, given the tracklets observed in a small window sampled on the surface. In our study, the particles are assumed to obey a stochastic Brownian motion with drift and may appear or disappear during the observation period. Split or merge events are not considered in the modeling framework. The trajectory reconstruction problem is defined as the maximization of the likelihood function given tracklets inside the ROO. The optimization problem to be solved is formulated as an integer linear programming problem. The final algorithm is a data-driven algorithm with no hidden parameter to be set by the user. We demonstrate the performance and robustness of our computational method on simulation data, by varying the ratio of observed to unobserved region, the drift and variance of particles, as well as the rates of birth and death of particles. 

The remainder of the paper is organized as follows. In Section 2, we present the problem formally and introduce notations. In Section 3, we describe the probabilistic framework, including Poisson processes used to describe birth and death events, and Brownian motion with drift to represent particle motion. We also describe the computational procedure aiming at connecting tracklets belonging to the same trajectory, and then recovering the dynamics of particles on the whole closed surface. Note that we suppose that the curvature of the cylinder is known and so that the movements are represented on a 2D unwrapped surface. In Section 4, the performance of our algorithm is evaluated on simulated data. Finally, we conclude and propose some future work. A summary of notations useful for the evaluation of the likelihood is given in Supplementary Materials (\ref{sm:notations}).

\section{Problem statement and notations}

We consider a probabilistic model to represent particles that are born, move and die on a cylindric membrane. Formally, let us denote $H$ and $ L$ the height and perimeter of the cylinder respectively (see Fig. \ref{fig:illus_partialobservation}). We associate 2D coordinates $(x,y) \in [-L,0] \times [0,H]$ to each point of the underlying cylindric manifold.  
The particles are "born" with a constant rate $\lambda$ and appear uniformly at random on the membrane surface. We consider a Poisson process with intensity $\lambda$  to statistically represent the birth events. Each particle is assumed to have the same constant rate of death $\tau_d$ such that life duration $T_d$ of a particle follows an exponential law of parameter $\tau_d$. 
During its lifetime, a particle $k$ born at time $t_0$ and located at $\mathbf{Z}_0^k=(X_0^k,Y_0^k)$, moves according to Brownian motion with drift. On the set $]-L,0[  \times [0,H],$ the position of the particle at time $t\geq t_0$ prior to its death time is given by
\begin{equation}
    \mathbf{Z}_t^k = \mathbf{Z}_0^k + \mathbf{v} (t-t_0) + \mathbf{\Sigma} \mathbf{B}_{t-t_0}^k
    \label{eq:bmd}
\end{equation}
where $\mathbf{Z}_t^k=(X_t^k,Y_t^k)$, $\mathbf{v}=(v_x,v_y)$, $\mathbf{\Sigma} =
  \left[ {\begin{array}{cc}
   \sigma_x & 0 \\
   0 & \sigma_y \\
  \end{array} } \right]$, $\mathbf{B}_t^k$ is a two-dimensional Wiener process. 
In order to model the topology of the cylinder as illustrated in Fig. \ref{fig:illus_partialobservation}, we impose deterministic jumps when the process reaches one of the two borders $\{-L\} \times [0,H]$ or $\{0\} \times [0,H]$. For any $y \in [0,H],$ the process reaching position $(-L,y)$ jumps to position $(0,y)$ and vice versa.  In $y$ direction the initial position of a particle lies between $[0, H]$. When a particle hits the vertical borders, its following trajectory is no longer considered. Finally, we assume that each particle behaves independently from the others and that there is no fission or fusion of particles. 

In the sequel, we observe the dynamics at discrete times $\Delta t, 2\Delta t, 3\Delta t \dots$ We denote $\Delta t$ the time step on the subset $[-l,0] \times [0,H]$ with $l<L.$ The observations are recorded during a time interval $[0,T_S]$. As we suppose that a particle does not change its drift direction along its trajectory, we assume that ${v}_x > 0$, even though particles can actually move in both directions, which requires a classification to separate them into two groups.  We consider that an observed tracklet of a given trajectory is an output if the last observed point of the segment is within a neighborhood of $\{0\} \times [0,H]$. Meanwhile, we consider that it is an input if the first observed point is within a neighborhood of $\{-l\}\times [0,H].$ 
Our main objective is then to associate the set of tracklets exiting the observed set $[-l,0] \times [0,H]$ with the set of tracklets entering this observation set. The challenge is to correctly match the outputs and the inputs associated to particles (see Fig. \ref{fig:illus_partialobservation}).

\section{Probabilistic models and methods}
Let us consider a given sample $S$, the observation set of all the trajectories. Define the sets $O_S=\{o_1,...,o_p\}$ and $I_S=\{i_1,...,i_q\}$ of $p$ outputs and $q$ inputs. Each output $o=(t_o,y_o) \in O_S$ is characterized by its output time $t_o$ and its position $y_o \in [0,H]$ where the particle left the observed region. Similarly each input $i=(t_i,y_i) \in I_S$ is characterized by its input time $t_i$ and its position $y_i \in [0,H]$ where it entered the observed region. A particle "involved" in an output $o \in O_S$ either died after time $t_o$ in the unobserved region, or is "involved" in a given input $i \in I_S$ with $t_i>t_o.$ We will denote this event by $\{o \to i\}.$ Similarly a particle "involved" in an input $i \in I_S$ was either born before time $t_i$ in the unobserved region, or is "involved" in a given output $o \in O_S$ with $t_i>t_o$, which corresponds to the event $\{o \to i\}.$

\begin{figure}[h]
\centering
\begin{tikzpicture}
[node distance = 1cm, auto,font=\footnotesize,
every node/.style={node distance=3cm},
comment/.style={rectangle, inner sep= 5pt, text width=3.7cm, node distance=0.25cm, font=\scriptsize\sffamily},
force/.style={rectangle, draw, fill=black!10, inner sep=5pt, text width=3.7cm, text badly centered, minimum height=1.2cm, font=\bfseries\scriptsize\sffamily}] 

\node [force] (datas) {Sample $S$, the outputs and inputs sets $O_S$ and $I_S$,\\ (section 3.1)};
\node [force, below = 1cm of datas] (estim) {Estimation of parameters $\hat{\tau}_\alpha, \hat{\tau}_d, \hat{v}$ and $\hat{\sigma}$,\\ (sections 3.1, 3.3)};
\node [force, below=2.5cm of estim, text width=4cm] (Proba) {Computation of the probability $P(\mathcal{B}_c)$ that the inputs in $B_c$ are born in the unobserved region. \\ \scriptsize{(section 3.1 Eqs. (\ref{eq:BcEps}) and (\ref{eq:ProbaBcEps}))}};
\node [force, left=0.1cm of Proba] (Probd) {Computation of the probability $P(D_c)$ that the outputs in $D_c$ died in the unobserved region.\\
\scriptsize{(section 3.1 Eq. (\ref{eq:death}))}};
\node [force, right=0.1cm of Proba] (Probc) {Computation of the probability $P(\flat_c)$ that a set of outputs are connected to a set of inputs.\\
\scriptsize{(section 3.1 Eqs. (\ref{eq:ProbabcEps}) and (\ref{eq:probaOtoI})}};
\node [force, below of=Proba, text width=4.5cm] (MaxiLikeli) {Likelihood\\ 
(section 3.1 Eqs. (\ref{eq:decomp1}), (\ref{eq=probaCepsfinal})-(\ref{eq:hatc}))};
\node [force, below  = 1cm of MaxiLikeli, text width=4cm] (cplex) {Maximization of likelihood with CPLEX, \\(section 3.2 Eqs. (\ref{eq:DecompArgMax}) and(\ref{eq:optim}))};
\node [force, below=1cm of cplex, text width=4cm] (criteriaAri) {Tracks reconstruction and estimation of connection accuracy, \\
(section 4.2)};


\node [comment, text width=5cm, below=0.1 of estim] (comment-Proba) {(+) Lifetime of a particle $T_d \sim \mathcal{E} (\tau_d)$\\
(+) The first passage time on $l$ of a Brownian motion $T_l$ follows an Inverse Gaussian distribution (Prop 1).};
\node [comment, right=0.25cm of estim] (comment-estim) {(+) To emphasize, $\hat{\tau}_\alpha$ is a novel estimator Eq.(19)};
\node [comment, right = 0.1cm of cplex] (comment-cplex) {(+) The possibility to find the n\textsuperscript{th} optimal solution, second part of section (3.2)};

\path[->,thick] 
(datas) edge (estim)
(comment-Proba) edge (Proba)
(comment-Proba) edge (Probd)
(comment-Proba) edge (Probc)
(Proba) edge (MaxiLikeli)
(Probd) edge (MaxiLikeli)
(Probc) edge (MaxiLikeli)
(MaxiLikeli) edge (cplex)
(cplex) edge (criteriaAri);

\end{tikzpicture} 
\caption{An outline of the connection procedure: from the estimation of the parameters to connection accuracy measurement, including likelihood formulation.  All notations are defined in the corresponding sections.}
\label{fig:procedure}
\end{figure}
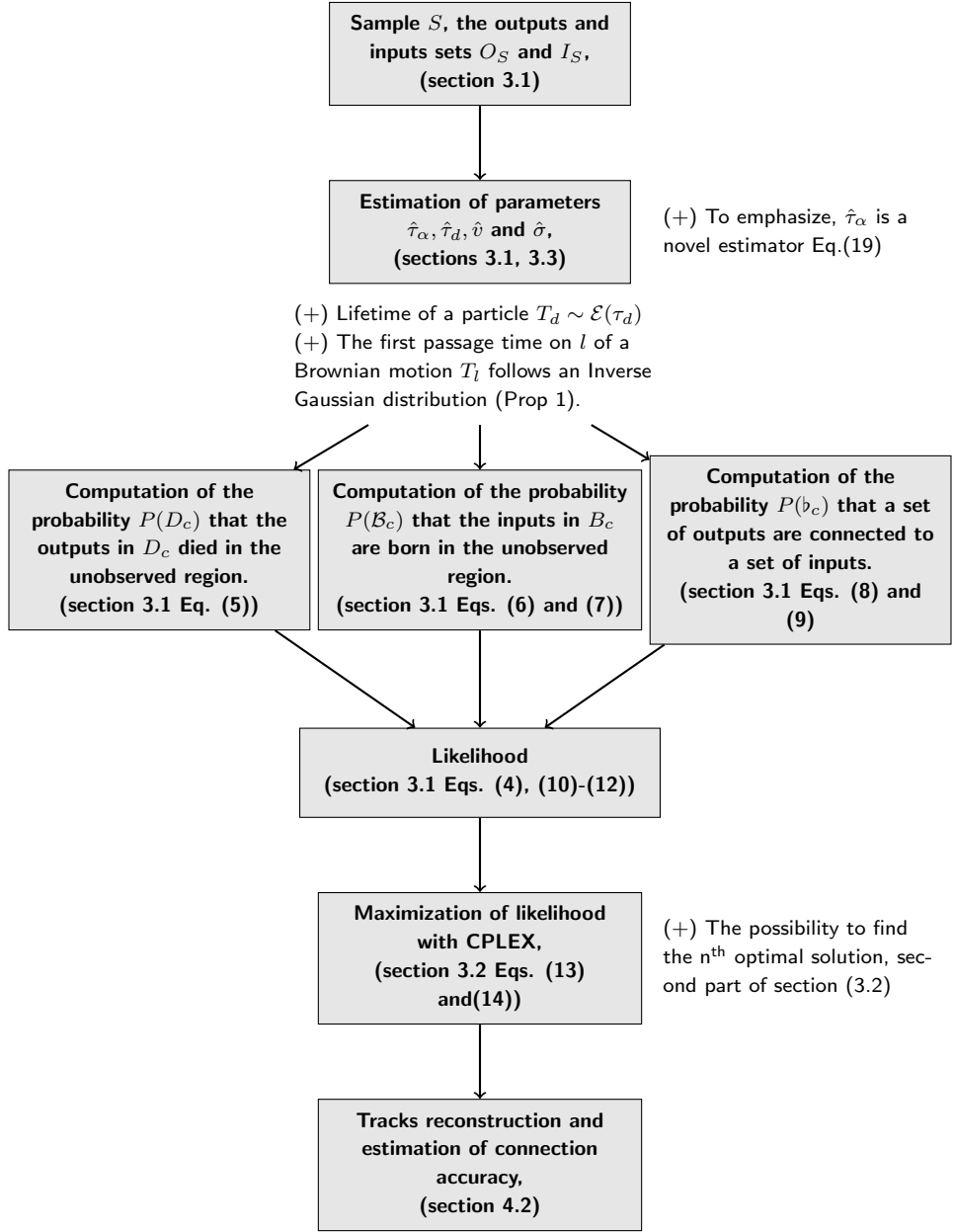

Define $c=(D_c,B_c,b_c)$ with $D_c \subset O_S, \, B_c \subset I_S$ and $b_c$ a bijection from $O_S \setminus D_c$ to $I_S \setminus B_c$ in order to describe the configuration for which all outputs in $D_c$ died in the unobserved region, all inputs in $B_c$ are born in the unobserved region, and the event $$\bigcap_{o \in O_s \setminus D_c} \{o \to b_c(o)\}$$ was realized.
Our aim is to determine the maximum likelihood configuration $c$ given the sample $S$. The outline of the connection procedure is given in Fig. \ref{fig:procedure}, to facilitate the understanding of the modeling steps.

\subsection{Likelihood of a configuration}\label{sec:likelihood}
In this section, our objective is to derive an analytic  expression of the likelihood $Q(c)$ of a configuration $c$. The aim is to find, for a given sample $S,$ the configuration $\hat c$ such that $P(\hat c / S)$ is maximal. It is difficult to calculate directly $P(\hat c / S)$. Since $c \subset S \subset O_S,$ we can compute $P(\hat c / S)$ working conditionally on $O_S.$ 

However, since the model is in continuous time and involves random variables with continuous densities with respect to the Lebesgue measure, the conditional probability $P(c / O_S )$ is equal to $0$.
%
This prevents to compute directly $P(\hat c / S)$ with the classical conditional formula 
$$P\left( c/S \right)=\frac{P\left( c / O_S\right)}{P(S/O_S)},$$
 because it gives $P(S/O_S)=\sum_{c \in \mathcal{C}_S} P(c / O_S) =0.$

Therefore, for each input $i=(t_i,y_i) \in I_S$, we consider  a spatiotemporal neighborhood $V_i^\epsilon = T_i^\epsilon \times H_i^\epsilon$ with $T_i^\epsilon= [t_i- \frac{\epsilon}{2}, t_i+ \frac{\epsilon}{2}]$ and $H_i^\epsilon= [y_i- \frac{\epsilon}{2}, y_i+ \frac{\epsilon}{2}]$ for some $\epsilon > 0.$ 

The idea is to replace a given configuration $c$ by a set $\mathcal{C}^\epsilon_c$ of configurations where each element $c^* \in \mathcal{C}^\epsilon_c$ is similar to $c$ but each input $i \in I_S$ is replaced by an input in $V_i^\epsilon.$ Formally, for each configuration $c$ leading to the input set $I_S,$  $\mathcal{C}^\epsilon_c$ is the set of configurations defined as follows:
 $c^*=(D_{c^*},B_{c^*},b_{c^*}) \in \mathcal{C}^\epsilon_c$ if and only if for each $i \in I_S,$ there exist $i_\epsilon^* \in V_i^\epsilon$ satisfying:
\begin{eqnarray*}
\left\{
\begin{array}{l}
D_{c^*}=D_c,\\
B_{c^*}= \{i_\epsilon^*, i \in B_c \},\\
\mbox{For each}~ i \in I_S \setminus B_c, \, b_{c^*}\left( b_c^{-1} (i) \right) = i_\epsilon^*.
\end{array}
\right.
\end{eqnarray*}
With this definition, we have

\begin{equation}\label{eq:LimEps}
P(c/S)= \lim_{\epsilon \to 0} P \left( \mathcal{C}^\epsilon_c / S \right) = \lim_{\epsilon \to 0} \frac{P \left( \mathcal{C}^\epsilon_c / O_S \right)}{\sum_{c'\in \mathcal{C}_S} P \left( \mathcal{C}^\epsilon_{c'} / O_S \right)}.
\end{equation}

In what follows,  we  study the behavior of $P(\mathcal{C}^\epsilon_c / O_S)$ when $\epsilon$ goes to $0.$ We will always work conditionally on the realization of the output set $O_S$ but we will keep this conditioning implicit and write $P(\mathcal{C}^\epsilon_c)$ instead of $P(\mathcal{C}^\epsilon_c / O_S)$ in order to simplify the notations.
The study of $P(\mathcal{C}^\epsilon_c)$ will involve the probability for a particle to die in the unobserved region but also the probability that a particle born in this unobserved region enters the observed one in a given spatiotemporal neighborhood $V_i^\epsilon.$ 

Furthermore, we assume that the particles born in the unobserved region, enter the observed one with a constant rate $\tau_\alpha$ and with a uniform distribution on $\{-l\} \times [0,H].$ This is consistent with the fact that the particles are born with constant rate $\lambda$ and appear uniformly at random on the membrane surface. Therefore, denote by $N_\alpha$ the Poisson process of intensity $\tau_\alpha$ counting the number of inputs involved by particles born in the unobserved region. 

Consider an output $o \in O_S$ and the possibility for the particle involved in $o$ to die in the unobserved region. We have the following proposition (see \cite{schrodinger1915}, \cite{tweedie1945inverse}, \cite{wald1973sequential}).

\begin{prop}\label{prop:IG}
Given the particle motion model as Brownian motion with drift as described in equation \ref{eq:bmd}, the first passage time noted as $T_l$ on the entrance line $\{-l\} \times [0,H]$ of a particle starting at position  $z_0=(0,y_0)$ for some $y_0 \in [0,H],$  follows a law of inverse Gaussian, that is $T_l \sim IG \left( \frac{l_u}{\mathbf{v}_x}, \big(\frac{l_u}{\sigma_x}\big)^2 \right)$ where $l_u:=L-l$ is the length of the unobserved region.  

Recall that if $X\sim IG(\mu,\lambda)$, then $X \geq 0$ almost surely, and for each $x\geq 0,$
\begin{equation}
    P(X\leq x)=\int_0^x\sqrt{\frac{\lambda}{2\pi y^3}}\exp \big(-\frac{\lambda(y-\mu)^2}{2\mu^2y}\big) dy.
\end{equation}

\end{prop}
In our framework, the event corresponding to the death of a particle with life duration $T_d$ following an exponential law of parameter $\tau_d$ in the unobserved region is precisely $\{T_d < T_l\}$.  
Hence, we can derive an explicit expression of $P(\mathcal{C}^\epsilon_c).$

Assume $\epsilon$ small enough so that for each $i,i' \in I_S, \, T_i^\epsilon \cap T_{i'}^\epsilon = \emptyset.$
For a given configuration $c$ and a given $\epsilon >0,$ we will write $\mathcal{C}^\epsilon_c = (D_c, \mathcal{B}_c^\epsilon, \flat_c^\epsilon)$ with $\mathcal{B}_c^\epsilon = \{B_{c^*}, c^* \in \mathcal{C}^\epsilon_c \}$ and $\flat_c^\epsilon = \{b_{c^*}, c^* \in \mathcal{C}^\epsilon_c \}.$

Due to the independent behavior of the particles, we have the following decomposition:
\begin{equation}\label{eq:decomp1}
P(\mathcal{C}^\epsilon_c)=P(D_c) P(\mathcal{B}_c^\epsilon) P(\flat_c^\epsilon).
\end{equation}
We can then compute separately the probabilities of events $D_c, \, \mathcal{B}_c^\epsilon$ and $\flat_c^\epsilon$.  First, note that we can assume without loss of generality that each output $o \in D_c$ starts at time $t_o=0$ and that only the position $y_o \in [0,H]$ fluctuates with $o$, but with no influence on $T_d$ or $T_l.$ Moreover, the loss of memory property of the exponential law ensures that the life duration $T_d$ of the particle after the output $o$ still follows an exponential law of parameter $\tau_d.$

Since all outputs behave identically and independently, we have $P(D_c) = P( T_d < T_l)^{|D_c|},$where $|D_c|$ stands for the cardinal of $D_c.$
According to proposition \ref{prop:IG}, and since $T_d$ and $T_l$ are independent, we have
\begin{eqnarray}\label{eq:death}
P( T_d < T_l) &=& \int_0^{+\infty} \int_0^{t_l} f_{T_d}(t_d) f_{T_l}(t_l) dt_d \; dt_l, \\
&=& \int_0^{+\infty} \int_0^{t_l} \tau_d e^{-\tau_d t_d} \frac{l_u}{\sigma_x\sqrt{2\pi t_l^3}}\exp \left(-\frac{(\mathbf{v}_xt_l-l_u)^2}{2\sigma_x^2t_l}\right) dt_d \;dt_l, \nonumber \\
&=& \int_0^{+\infty}  \frac{l_u \left( 1 - e^{-\tau_d t_l} \right)}{\sigma_x\sqrt{2\pi t_l^3}}\exp \left(-\frac{(\mathbf{v}_xt_l-l_u)^2}{2\sigma_x^2t_l}\right) dt_l, \nonumber 
\end{eqnarray}
where $f_{T_d}$ and $f_{T_l}$ respectively stand for the density functions of $T_d$ and $T_l.$

Now, consider the event $\mathcal{B}_c^\epsilon.$ We call "spontaneous input" an input related to a particle born in the unobserved region that has never been observed. The set $\mathcal{B}_c^\epsilon$ is defined so that, for each input $i \in B_c$, we have exactly one "spontaneous input" appearing during the time interval $T_i^\epsilon,$ with a position in $H_i^\epsilon.$ Moreover, outside $\cup_{i \in B_c} T_i^\epsilon,$ there is no "spontaneous input". Formally, we have
\begin{equation}\label{eq:BcEps}
\mathcal{B}_c^\epsilon = \left\{ N_\alpha \left( [0,T_S] \setminus \bigcup_{i \in B_c} T_i^\epsilon \right) =0 \right\} \bigcap \left( \bigcap_{i \in B_c} \Big( \{N_\alpha (T_i^\epsilon)=1\} \cap H_i^\epsilon \Big) \right),
\end{equation}
where $N_\alpha$ is a Poisson process of intensity $\tau_\alpha$ associated to the counting of inputs involved by particles born in the unobserved region on the time interval $[0,T_S].$
In order to simplify the notations, $H_i^\epsilon$ denotes also the event of "spontaneous" appearance of an input $i$ in $H_i^\epsilon$. This event is independent of the process $N_\alpha$, and since the "spontaneous inputs" appear uniformly on $[0,H],$ we have $P(H_i^\epsilon)=\frac{\epsilon}{H}.$

Meanwhile, for any time interval $I, \, N_\alpha(I)$ follows a Poisson law of parameter $\tau_\alpha |I|$ where $|I|$ denotes the length of the interval $I.$ Since $\epsilon$ is small enough so that for each $i,i' \in I_S, \, T_i^\epsilon \cap T_{i'}^\epsilon = \emptyset, \, N_\alpha( T_i^\epsilon)$ and  $N_\alpha( T_{i'}^\epsilon)$ are independent.
Consequently, we can compute $P(\mathcal{B}_c^\epsilon)$ as follows: 
\begin{eqnarray}\label{eq:ProbaBcEps}
P(\mathcal{B}_c^\epsilon)  = e^{-\tau_\alpha (T_S - |B_c|\epsilon)} \left( \epsilon \tau_\alpha e^{-\epsilon \tau_\alpha} \frac{\epsilon}{H} \right)^{|B_c|} 
& = \left( \frac{\epsilon^2 \tau_\alpha}{H} \right)^{|B_c|} e^{-\tau_\alpha T_S}.
\end{eqnarray}

Finally, consider the event $\flat_c^\epsilon$. For each input $i \in I_S \setminus B_c,$ we denote by $\{ o_c^i \to V_i^\epsilon \}$ the survival event of  the particle involved in the output $o_c^i=b_c^{-1}(i)$ in the unobserved region which appears in the spatiotemporal neighborhood $V_i^\epsilon$. Since the particles behave independently, we have
\begin{equation}
\label{eq:ProbabcEps}
P(\flat_c^\epsilon)=\prod_{i \in I_S \setminus B_c} P \Big(\{ o_c^i \to V_i^\epsilon \} \Big).
\end{equation}

In the sequel, we consider a given input $i \in I_S \setminus B_c$ and its related output $o=b_c^{-1}(i).$ Defining $s_i=t_i-t_o$ and $h_i=y_i-y_o$ allows us to center the situation around the output $o$ in the following way. 
A particle born at time $0$ in position $z_0=(0,0)$ has a life duration $T_d$ following an exponential law of parameter $\tau_d.$ During its lifetime, the position of the particle is driven  by a Brownian motion with drift $\mathbf{Z}_t=(X_t,Y_t)$: $\mathbf{Z}_t = \mathbf{v} t + \mathbf{\Sigma} \mathbf{B}_t,$ where $\mathbf{B}_t$ is a two-dimensional Wiener process and $\mathbf{v}$ and $\mathbf{\Sigma}$ are given  in Equation (\ref{eq:bmd}). 
Define $T_{l}$ the first reaching time of $l_u=L-l$ of the process $X_t$. The event $\{ o \to V_i^\epsilon \}$ can now be written as follows:
\begin{equation*}
\!\!\!\!\!\!\!\!\!\!\!\!\!\!\!\{ o_c^i \to V_i^\epsilon \}= \left\{T_d > T_l \right\} \bigcap \left\{T_l \in \left[s_i- \frac{\epsilon}{2},s_i + \frac{\epsilon}{2}\right] \right\} \bigcap \left\{Y_{T_l} \in \left[h_i- \frac{\epsilon}{2},h_i + \frac{\epsilon}{2}\right] \right\}.
\end{equation*}
This expression corresponds exactly to the fact that in order to realize $\{ o_c^i \to V_i^\epsilon \}$ the particle needs to have a life duration longer than its first reaching time of $l_u$ and to appear in the spatiotemporal neighborhood $ \left[s_i- \frac{\epsilon}{2},s_i + \frac{\epsilon}{2}\right] \times \left[h_i- \frac{\epsilon}{2},h_i + \frac{\epsilon}{2}\right].$
Furthermore, $T_d$ follows an exponential law of parameter $\tau_d,$ $Y_t$ follows a Gaussian law of parameters $\mathbf{v}_yt$ and $\sigma_y^2t$ and  $T_l \sim IG \left( \frac{l_u}{\mathbf{v}_x}, \big(\frac{l_u}{\sigma_x}\big)^2 \right)$.  Moreover, due to the fact that $\mathbf{\Sigma}$ is diagonal, the process $Y_t$ is not only independent of $T_d$ but also of $T_l.$ This allows us to write
$$
P \left( \{ o_c^i \to V_i^\epsilon \} \right)= \int_{s_i-\frac{\epsilon}{2}}^{s_i + \frac{\epsilon}{2}} f_{T_l}(t_l) \left( \int_{t_l}^{+ \infty} f_{T_d}(t_d) \left( \int_{h_i-\frac{\epsilon}{2}}^{h_i + \frac{\epsilon}{2}} f_{Y_{t_l}}(y) dy \right) dt_d \right) dt_l.
$$
As the two integrals involve a small domain of size $\epsilon, \, P \left( \{ o_c^i \to V_i^\epsilon \} \right) \sim \mathcal{O} (\epsilon^2)$, and 
\begin{eqnarray}\label{eq:probaOtoI}
\!\!\!\!\!\!\!\!\!\!\!\!\!\!\!\!\!\!\!\!\!\!\!\!\!\!\!\!\!\!\!\!\!\! \lim_{\epsilon \to 0} \frac{P \left( \{ o_c^i \to V_i^\epsilon \} \right)}{\epsilon^2} = f_{T_l}(s_i) f_{Y_{s_i}}(h_i) \int_{s_i}^{+ \infty} f_{T_d}(u) du  \\
=  \frac{l_u}{\sigma_x\sqrt{2\pi s_i^3}}\exp \left(-\frac{(\mathbf{v}_xs_i-l_u)^2}{2\sigma_x^2s_i}\right) \frac{1}{\sigma_y \sqrt{ 2 \pi s_i}} \exp \left( - \frac{\left(h_i -  \mathbf{v}_y s_i\right)^2 }{2 \sigma_y^2 s_i} \right) e^{-\tau_d s_i} \nonumber \\
= \frac{l_u}{2 \pi \sigma_x \sigma_y s_i^2} \exp \left( -\frac{(\mathbf{v}_xs_i-l_u)^2}{2\sigma_x^2s_i} - \frac{\left(h_i -  \mathbf{v}_y s_i\right)^2 }{2 \sigma_y^2 s_i} -\tau_d s_i \right). \nonumber
\end{eqnarray}

For each configuration $c$, we calculate the likelihood $Q(c)$ of the configuration $c$ as follows:
$$Q(c)~:= ~\lim_{\epsilon \to 0} \frac{P(\mathcal{C}^\epsilon_c)}{\epsilon^{2|I_S|}}.$$
From (\ref{eq:decomp1}) and Equations (\ref{eq:death}, \ref{eq:ProbaBcEps}, \ref{eq:ProbabcEps} and \ref{eq:probaOtoI}), we finally obtain the likelihood 
\begin{eqnarray}\label{eq=probaCepsfinal}
\! \! \! \! \! \!\! \! \! \! \! \! \! \! \!\! \! \!\! \! \! \!  Q(c) ~=~\left( \frac{\tau_\alpha}{H} \right)^{|B_c|} e^{-\tau_\alpha T_S} \left(\int_0^{+\infty}  \frac{l_u \left( 1 - e^{-\tau_d t_l} \right)}{\sigma_x\sqrt{2\pi t_l^3}}\exp \left(-\frac{(\mathbf{v}_xt_l-l_u)^2}{2\sigma_x^2t_l}\right) dt_l \right)^{|D_c|} \nonumber \\
~~~~\times \prod_{i \in I_S \setminus B_c} \left[ \frac{l_u}{2 \pi \sigma_x \sigma_y s_i^2} \exp \left( -\frac{(\mathbf{v}_xs_i-l_u)^2}{2\sigma_x^2s_i} - \frac{\left(h_i -  \mathbf{v}_y s_i\right)^2 }{2 \sigma_y^2 s_i} -\tau_d s_i \right) \right].
\end{eqnarray}

Note that the limit when $\epsilon$ goes to $0$ of $\frac{P(\mathcal{C}^\epsilon_c)}{\epsilon^{2|I_S|}}$ is well defined, strictly positive, and that the exponent $2 |I_S|$ does not depend on the configuration $c.$

Recalling (\ref{eq:LimEps}), this allows us to write

\begin{equation}\label{eq:LimEpsQ}
P(c/S)=  \frac{Q(c)}{\sum_{c'\in \mathcal{C}_S} Q(c')}
\end{equation}

and as a consequence, we have 

\begin{equation}\label{eq:hatc}
\hat c= \arg\!\max_{c \in \mathcal{C}_S} \{ Q(c) \}.
\end{equation}


\subsection{Maximum likelihood and optimal configuration} 

The aim of this section is to identify the configuration $c$ corresponding to the maximal likelihood $Q(c)$ (see Equation (\ref{eq=probaCepsfinal})). Define 
$$\beta:= - \log \left(\frac{\tau_\alpha}{H}\right),$$
$$
\delta := - \log \left(\int_0^{+\infty}  \frac{l_u \left( 1 - e^{-\tau_d t_l} \right)}{\sigma_x\sqrt{2\pi t_l^3}}\exp \left(-\frac{(\mathbf{v}_xt_l-l_u)^2}{2\sigma_x^2t_l}\right) dt_l \right)
$$
and for each configuration $c$ and each $i \in I_S \setminus B_c$
$$\gamma_c^i:= - \log \left[ \frac{l_u}{2 \pi \sigma_x \sigma_y s_i^2} \exp \left( -\frac{(\mathbf{v}_xs_i-l_u)^2}{2\sigma_x^2s_i} - \frac{\left(h_i -  \mathbf{v}_y s_i\right)^2 }{2 \sigma_y^2 s_i} -\tau_d s_i \right) \right].$$
It follows that 
\begin{eqnarray}\label{eq:DecompArgMax}
\hat c ~=~  \arg\!\max_{c\in C} Q(c) &=& \arg\!\min_{c\in C} - \log \left(Q(c) \right) \\\nonumber
 &=& \arg\!\min_{c\in C} \left( \beta |B_c| + \delta |D_c| + \sum_{i \in I_S \setminus B_c} \gamma_c^i \right).
\end{eqnarray}
This decomposition allows us to consider a linear optimization problem where $\beta$ represents the cost of the spontaneous birth of an input, $\delta$ the cost of the death of an output and $\gamma_c^i$ the cost of the connection between the output $b_c^{-1}(i)$ and the input $i.$
The cost of connection can be defined for any couple $(o,i) \in O_S \times I_S$ as
$$\gamma_o^i:= - \log \left[ \frac{l_u}{2 \pi \sigma_x \sigma_y s_{o , i}^2} \exp \left( -\frac{(\mathbf{v}_xs_{o , i}-l_u)^2}{2\sigma_x^2s_{o , i}} - \frac{\left(h_{o , i} -  \mathbf{v}_y s_{o , i}\right)^2 }{2 \sigma_y^2 s_{o , i}} -\tau_d s_{o , i} \right) \right],$$
where $s_{o,i}:= t_i-t_o, \, h_{o,i}=y_i-y_o$ and the convention $\gamma_o^i = + \infty$ if $t_i \leq t_o.$

In order to write in a canonical way this linear optimization problem, we associate to each configuration $c$ a family of coefficients $ (c^{o,i})_{(o,i) \in O_S \times I_S}$ such that $c^{o,i}=1$ if $b_c(o)=i$ and $c^{o,i}=0$ if $b_c(o) \neq i.$ 
Since an output can be connected to at most one input, for each $o \in O_S$,  $\sum_{i \in I_S}c^{o,i} \in \{0,1\}$ and $\sum_{i \in I_S}c^{o,i} = 0$ corresponds to the death of the output $o.$ Similarly, for each $i \in I_S,$ $\sum_{o \in O_S}c^{o,i} \in \{0,1\}$ and $\sum_{o \in O_S}c^{o,i} = 0$ corresponds to the fact that the input $i$ is a "spontaneous input".

Our optimization problem is then equivalent to finding the family of coefficients $ (c^{o,i})_{(o,i) \in O_S \times I_S}$ that minimizes the quantity
$$
\beta \left(\sum_{i \in I_S} \left(1- \sum_{o \in O_S} c^{o,i} \right) \right) + \delta \left( \sum_{o \in O_S} \left(1- \sum_{i \in I_S} c^{o,i} \right) \right) + \sum_{o \in O_S} \sum_{i \in I_S} \gamma_o^i c^{o,i}
$$
or equivalently 
\begin{eqnarray}
~~~~~~K(c):=\sum_{o \in O_S} \sum_{i \in I_S} \left(\gamma_o^i -\beta -\delta \right) c^{o,i}   ~\mbox{s.t.}~
\left\{
\begin{array}{l}
\forall o \in O_S, \, \forall i \in I_S, \, c^{o,i} \in \{0,1\}, \\ 
\forall o \in O_S, \, \sum_{i \in I_S} c^{o,i} \in \{0,1\},\\ 
\forall i \in I_S, \, \sum_{o \in O_S} c^{o,i} \in \{0,1\}. 
\end{array}
\right.
\label{eq:optim}
\end{eqnarray}
In order to avoid to have infinite costs $\gamma_o^i$ when $t_i \leq t_o,$, we can also impose  $c^{o,i}=0$ if $t_i \leq t_o$. Actually the problem (\ref{eq:optim}) is a conventional linear optimization problem which  can be solved by applying the CPLEX Linear Programming solver (https://www.ibm.com/analytics/cplex-optimizer).

The configuration $\hat c$ is then the solution of the optimization problem (\ref{eq:optim}) and corresponds to the most likely configuration given the sample $S.$ In order to complete the study, we propose to compute the following most likely configurations in a reccurent way by solving (\ref{eq:optim}) with additional constraints ensuring that the solution is different from the previous ones. In other words we define recursively the sequence $\left( c_n \right)_{n \in \mathbb N}$ in the following way:

\begin{itemize}
\item $c_1 := \hat c$
\item $\forall n \geq 2, \, c_n$ solves (\ref{eq:optim}) with the $n-1$ additional constraints 
\begin{equation}\label{neme_sol}
\forall k  \in \{ 1, \dots, n-1 \}, \,  \sum_{o \in O_S} \sum_{i \in I_S} \left[ c_n^{o,i} (1-c_k^{o,i}) + (1-c_n^{o,i})c_k^{o,i} \right] \geq 1.
\end{equation}
\end{itemize}

With this definition, $c_n$ is then the $n-$th most likely configuration. When $n$ is greater than the number $n_S$ of configurations compatible with the sample $S,$ the constraints are impossible to satisfy. In other words this sequence is well defined up to $n_S.$

\subsection{Estimation of parameters}\label{sec:estimators}

Several parameters are involved in our computational approach. In this section, we propose clues to set these parameters.   
First, the parameters $\mathbf{v}$ and $\mathbf{\Sigma}$ can be estimated with classical maximum likelihood estimation procedures.

Second, we propose an estimator $\hat \tau_d$ of $\tau_d$ as explained below.
The sample $S$ can be considered as a set of points $p=(t_p,\mathbf{Z}_p)$ observed at time $t_p$ and position $\mathbf{Z}_p=(X_p,Y_p)$  grouped in clusters $s$ corresponding to tracklets of trajectories.
The death of a particle in the observed region is detected in $S$ for each point $p \in S$ for which the associated tracklet $s_p$ has no successor point at time $t_p+\Delta t$. In order to be sure that the absence of successor is effectively due to the death of a particle and not to a particle leaving the observed region, we restrict the analysis to a region excluding a neighborhood of the border. However,  we can check in this neighborhood the existence of successors for points in the restricted region. We denote by $S_r \subset S$ the sample of points in the restricted region. 
For each point $p \in S_r,$ we denote by $D_p$ the event corresponding to the absence of successor for $p.$ This corresponds to the fact that the particle involved in $p$ died during the time interval $[t_p, t_p + \Delta t].$ Since the life duration $T_d$ of a particle follows an exponential law of parameter $\tau_d$, and the absence of memory property of the exponential law, we have 
\begin{equation}\label{eq:probDp}
P(D_p)=P(T_d \in [0,\Delta t])=1-e^{-\tau_d \Delta t}.
\end{equation}
Hence, we define our estimator $\hat \tau_d$ as
\begin{equation}\label{eq:defHatTaud}
\hat \tau_d= \frac{1}{\Delta t |S_r|} \sum_{p \in S_r} 1[D_p],
\end{equation}
where $|S_r|$ stands for the number of points in $S_r$ and $1[\cdot]$ denotes the indicator function.  
Due to the absence of memory property of the exponential law, the random variables $1[D_p]$ are i.i.d.  As $|S_r|$ goes to $+ \infty,$ the strong law of large numbers yields to 
$$
\lim_{|S_r| \to \infty} \hat \tau_d = \frac{1-e^{-\tau_d \Delta t}}{\Delta t} \quad \quad a.s.
$$
The justification of this choice for $\hat \tau_d$ relies in the following almost sure convergence:
\begin{equation}\label{eq:HatTauConsistent}
\lim_{\Delta t \to 0} \lim_{|S_r| \to \infty} \hat \tau_d = \tau_d \quad \quad a.s.
\end{equation}
Our estimator $\hat \tau_d$ is then consistent as $\Delta t$ is small enough.
Moreover, since the variables $1[D_p]$ are i.i.d Bernoulli random variables, we can calculate the related confidence interval. If $q_\alpha$ denotes the $\alpha$-quantile of the standard normal distribution, we have the following confidence interval of level $\alpha$ for  $\frac{1-e^{-\tau_d \Delta t}}{\Delta t}$:
\begin{equation}
CI_\alpha = \left[  \hat \tau_d - q_\alpha \sqrt{\frac{\hat \tau_d \left( \frac{1}{\Delta t} - \hat \tau_d \right)}{|S_r|}} ,  \hat \tau_d + q_\alpha \sqrt{\frac{\hat \tau_d \left( \frac{1}{\Delta t} - \hat \tau_d \right)}{|S_r|}} \right].
\end{equation}
If $\Delta t$ is small enough, we get a good approximation of a confidence interval of level $\alpha$ for $\tau_d$ since 
\begin{equation*}
\lim_{\Delta t \to 0} \frac{1-e^{-\tau_d \Delta t}}{\Delta t} = \tau_d.
\end{equation*}

Now, we describe the estimation procedure for the rate $\tau_\alpha$ of "spontaneous inputs" induced by particles born in the unobserved region $[-L,-l]\times [0,H]$ and reached the border $\{-l\}\times[0,H].$ We assume here that the parameters $\mathbf{v}, \, \mathbf{\Sigma}$ and $\tau_d$ are known, keeping in mind that in practice estimators are used instead. As introduced earlier, $L$ is the perimeter of the cylinder,  $l$ is the length of the observed region, and $l_u=L-l$ is the length of the unobserved region.  For a given length $x$, we denote by $N_x$ the number of tracklets born in the region $]-x,0]\times[0,H]$ and reached the border $\{0\}\times[0,H].$ Accordingly, $\frac{N_{l_u}}{T_S}$ is a consistent estimator of $\tau_\alpha$ since the dynamics are assumed to be homogeneous on the surface of cylinder. Our aim is actually to build an estimator for $\tau_\alpha$ in the case where $l_u >l$ which prevents us to compute directly $N_{l_u}$.
Therefore, we compute $N_l$ by taking the whole observed region into account, and denote by $S_l^*$ the set of tracklets having an input in $\{-l\}\times[0,H]$ and an output in $\{0\}\times[0,H].$  For each tracklet $s \in S_l^*$ and each length $x \in [0,l_u],$ we denote by $B_s^x$ the event corresponding to the birth of  $s$ within $]-l-x,-l]\times [0,H].$ Let $l_e=l_u-l$ be the length of the extended zone $[-l_u,-l] \times [0,H]$. We are now interested in the realization of the events $B_s^{l_e}.$

\begin{figure}[t]
    \centering
    \includegraphics[scale=0.4]{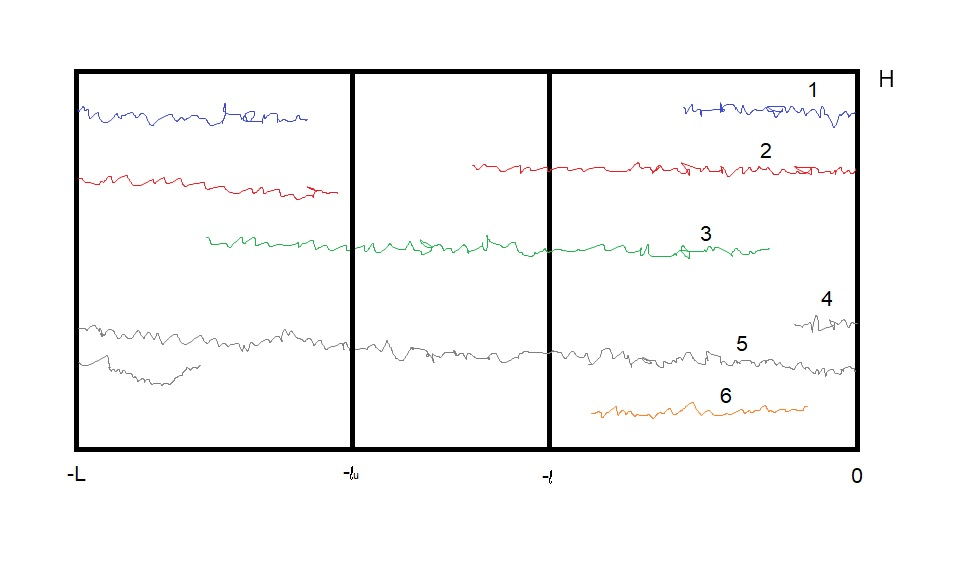}
    \caption{An artificially constructed zone $]-l_u,0]\times [0,H]$ having the same size as the unobserved region $]-L,-l]\times [0,H].$ The observed region is $]-l,0]\times [0,H]$ ; as the width of the invisible part is $l_u$, the extended zone has width $l_e=l_u-l.$}
    \label{fig:illust_Bs}
\end{figure}

In Fig. \ref{fig:illust_Bs}, $N_l=2$ correspond to tracks \#1 and \#4,  $S_l^*= \{2,5\},$ and the event $B_2^{l_e}$ is realized while $B_5^{l_e}$ is not.

Note that since the particles have the same independent dynamics, $P\left( B_s^x \right)$ does not depend on $s.$ For $x < l,$ this probability can easily be estimated as follows:
$$\hat p_x = \frac{N_x}{\big| S_o \big|},$$
where  $S_o$ is the set of tracklets having an output in $\{0\}\times[0,H].$ The strong law of large numbers yields a consistent estimator and allows us, in the case where $l_e <l,$ to define our estimator $\hat \tau_\alpha$ as follows:
\begin{equation}\label{eq:defHatTauAlpha}
\hat \tau_\alpha = \frac{N_l + \hat p_{l_e} |S_l^*|}{T_S}.
\end{equation}
Intuitively, this estimator amounts to counting the number of particles reaching $\{0\} \times [0,H]$ with weight 1 for each tracklet that we actually saw being born in the observed region and with weight $\hat p_{l_e}$ for each spontaneous input that appeared in $\{-l\} \times [0,H].$
Note that, as $N_{l_u}=N_l + \sum_{s \in S_l^*} 1[B_s^{l_e}], \, \hat \tau_\alpha$ is an unbiased estimator of $\tau_\alpha.$ Moreover, if we assume that the number of observed tracklets grows linearly with the observation time $T_S,$ this estimator is consistent when $T_S$ goes to $+ \infty.$

Now, we consider the case $l< l_e <2l$ which can easily be extended to the general case $l< l_e.$ 
Consider $s \in S_l^*$ and denote for each interval $J \subset [-L,0]$ the event $B_s^J$ where the tracklet $s$ is born in the region $J \times [0,H].$
The event $B_s^{l_e}$ can be decomposed as follows:
$$B_s^{l_e} = B_s^{[-2l,-l]} \bigcup \left( \overline{B_s^{[-2l,-l]}} \cap B_s^{[-l_u,-2l]} \right).$$
The loss of memory and homogeneity properties of the dynamics lead to the following  estimator $\hat p_{l_e}$:
$$\hat p_{l_e} := \hat p_l + \left( 1-\hat p_l \right) \hat p_{l_e-l}.$$

\subsection{Limits of the model}

The main assumptions in this work are homogeneity in time and space, induced by the constant death and birth rates, as well as constant speed and noise. While these assumptions lead to a simple model and allows a reasonably technical study, it is natural to question it. The main reason of this choice is that it corresponds to uniform laws when we have no reason prioritize one specific behavior in particular. 

Note that a similar study can be made with different speeds among trajectories. This can be done by classifying the trajectories according to their speeds and applying the present procedure to each class. This would lead to the same estimation procedure with smaller datasets but theoretical results will still hold.

We then discuss the homogeneity in time, for which the most questionable assumption is the constant death rate that could possibly depend on the position or on the age of the particle. Concerning the dependence in space, this modification would lead to the estimation of a function of the position instead of the simple constant $\tau_d.$ From a practical point of view, this would increase the dimension of the parameter to estimate, with the same size of dataset. From a theoretical point of view a more technical study can be made as long as we assume the death function rate (depending on the position) constant on each tracklet $\{y\}\times ]-L,0]$ in order to overcome the issue of partial observation.

Concerning the dependence in time, the assumption that the death rate depends on the age of the particle prevents to propose a similar study. Indeed, due to partial observation, the age of each particle entering the observed region is unknown and can not be estimated.

\subsection{Modeling hypothesis and MreB dynamics}
The study of the dynamics of MreB patches or assemblies in the vicinity of the internal membrane of \textit{Bacillus subtilis} bacteria reveals several subpopulations undergoing constrained, randomly or directionally moving \cite{billaudeau2017contrasting}. Herein we are interested in the directionally moving subpopulation dynamics. This subpopulation  moves possessively around the cell diameter \cite{garner2011coupled,dominguez2011processive}. Following Hussain \textit{et al} \cite{hussain2018mreb}, Billaudeau \textit{et al} \cite{billaudeau2019mreb} confirmed that directionally moving filaments travel in a direction close to their main axis, perpendicularly to the long axis of the cell (angle $\gamma=89.9^{\circ}\pm 37.0^{\circ}$). Hence, for some filaments, the speed vector may have a component in the main direction of the bacteria. 

According to Wong et al. \cite{wong2019mechanics} a motion model (named ``biased random walk'') reproduces the dynamics patterns of MreB filaments. In their simulations, the speed is constant, the noise variance between several time steps depends on the duration and, possibly on the local curvature of the surface. These properties are shared with the  Brownian motion model with constant drift we consider. 

\section{Simulation study}
In this section, we present a series of experiments performed on synthetic datasets. These experiments aim to evaluate and analyze the sensitivity of the reconstruction procedure when the characteristics of the dynamics as well as the spatio-temporal sampling resolution of observations vary. In addition to demonstrate the potential of our procedure, these experiments might also be useful for the design of the experimental setting for images acquisition. The reconstruction procedure has been implemented in \textit{MATLAB ver. R2018b}. The codes are available on Github \url{https://github.com/atrubuil/ReconstructionOfTruncatedTrajectories}.

\subsection{Generation of trajectories}
\begin{figure}[t]
\centering
\includegraphics[scale=0.7]{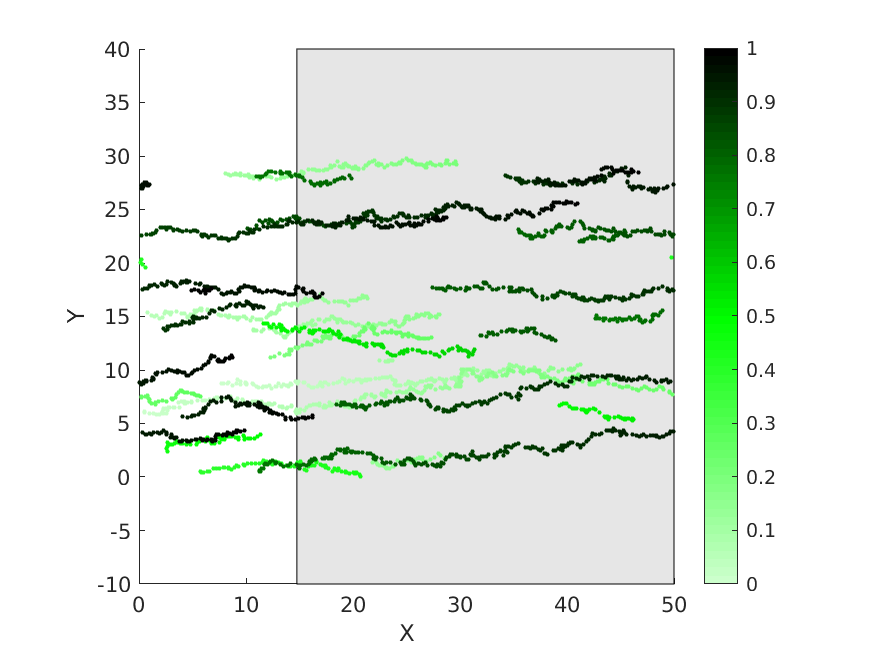}
\caption{A set of simulated trajectories during 2.5 minutes (in stationary regime). X(resp. Y) axis represents the unfolded circumferential (resp. main) direction of the cylinder. Colors from light to dark green represent time evolution. Shadowed area corresponds to the unobserved region and white area corresponds to the ROO.}
\label{tracks_gen}
\end{figure}

Trajectories are generated on a rectangular unwrapped cylindrical surface of size $[0,L]\times[0,H]$ (Fig. \ref{tracks_gen}).  In our experiments, we set $L = 50, H = 30$. The initial position of each trajectory is drawn from uniform distribution on the surface. Time duration $T$ between two births follows an exponential law with birth rate parameter $\lambda$. At each birth, the intrinsic properties of a trajectory $i$ are given, such as velocity $\mathbf{v}_i$, variance $\mathbf{\Sigma}_i$, and lifetime $T_d^i$. The lifetime $T_d$ follows an exponential law, with the same death rate $\tau_d$ for all trajectories in the whole simulated image sequence. The drift $\mathbf{v}_i = (v_{xi}, v_{yi})$ and noise $\mathbf{\Sigma}_i =
  \left[ {\begin{array}{cc}
   \sigma_{xi} & 0 \\
   0 & \sigma_{yi} \\
  \end{array} } \right]$ are set to be constant along one given trajectory. 
  
  According to the assumptions made on real biological context, unless otherwise stated, it is set by default, $\theta = 0.01 (\approx 0.6^{\circ})$ is the angle between the direction of motion of particle and the $X$ direction, $v_y = \tan(\theta)v_x$, $ \sigma_x = \sigma_y= \sigma$, $v_x = 0.6$, $\sigma = 0.2$, $\lambda = 0.03$, $\tau_d=0.005$. The time interval between two images $\Delta t = 0.25$. As known, the theoretical depth of the observation field of TIRFM is $200nm$, the diameter of the bacteria cell is $1\mu m$, therefore the width of the ROO $l$ is set to 14.76 and that of the unobserved region $l_u=35.24$ (unit in pixel, note that in TIRF images 1 pixel $\approx 64 nm$).

\begin{figure}[t]
\centering
\hspace{-0.5cm}\includegraphics[scale=0.6]{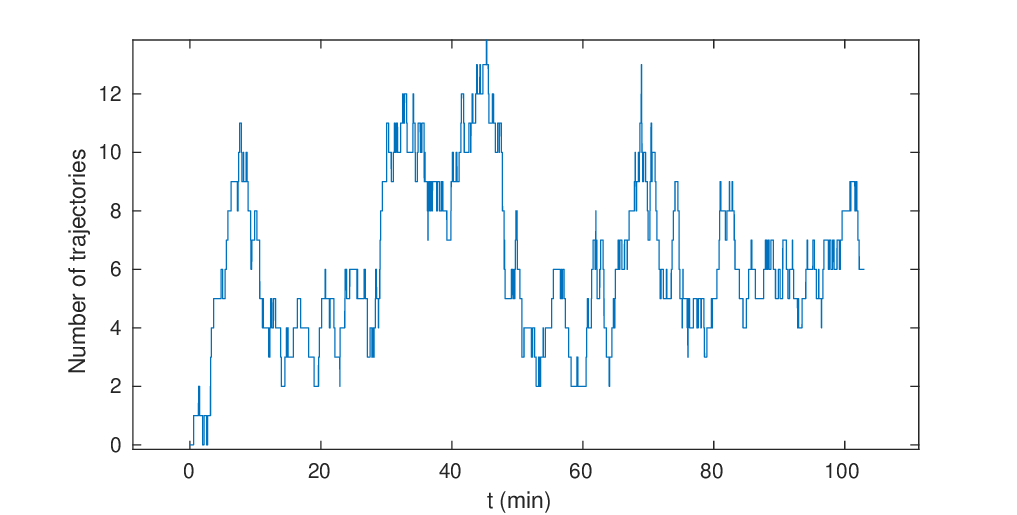}
\caption{Fluctuations of the number of trajectories w.r.t. time. At around $t = 20$ min, the trajectories number fluctuates around the theoretical expectation value 6.}
\label{nb_tracks}
\end{figure}

As there is no particle on the surface at the beginning, the simulated set of trajectories needs some warm-up time to reach the stationary regime, where the law of the number of trajectories does not depend on time. The assumed dynamic process is a special case of birth and death process.  As a known result\cite{karlin2014first}, the expectation of the trajectories number $N$ during stationary regime is $E(N)=\frac{\lambda}{\tau_d}$. To ensure that the dynamics are in a stationary regime, the images sequence is simulated long enough, for around 2 hours (Fig. \ref{nb_tracks}).

\subsection{The "Adjusted Rand Index" for the evaluation of connection results }

Given the true and estimated class assignments, we compute the so called Adjusted Rand Index  to evaluate similarity or consensus between the two sets. The Adjusted Rand index is the corrected-for-chance version of the Rand index. It is scored exactly 1 when the two sets are identical, close to 0 for random labeling. It could be negative when the index is lower than the expectation under random labeling. 
More precisely, let $G$ and $K$ be the true and estimated assignments respectively, let us define $a$ and $b$ as:
$a$ the number of pairs of elements that are in the same class in G and in the same class in K, $b$ the number of pairs of elements that are in different classes in G and in different classes in K. 
The raw (unadjusted) Rand index is then given by:
\begin{equation}
RI = \frac{a+b}{C_2^M},
\end{equation}
where $C_2^M$ is the total number of possible pairs in the dataset (without ordering) of size $M$.
The RI score does not guarantee that random assignments will get a value close to zero. This is especially true if the number of clusters has the same order of magnitude as the number of samples.
To overcome this difficulty, we prefer to consider the  Adjusted Rand Index defined by \cite{rand1971objective}:
\begin{equation}
ARI = \frac{RI-E(RI)}{1-E(RI)}.
\end{equation}

Here $E(RI)$ denotes the expectation of the Rand Index where the estimated assignment $K$ is replaced by an assignment chosen uniformly at random. This means that the assignment procedure does not do better than random assignment if the $ARI$ score is zero, and that it does worse than random if $ARI < 0.$

\subsection{Experimental results}
The good performance of the connection procedure relies on the estimation of the characteristics of the dynamics: the speed, $\mathbf{v}$, the diffusion variance, $\mathbf{\Sigma}$, the arrival rate $\tau_{\alpha}$ and the death rate $\tau_d$, as these quantities are used in the calculation of the likelihood (Eq. \ref{eq:DecompArgMax}). Here we evaluated the impact of spatio-temporal sampling $(l/l_u, T_S)$ on the estimators and the impact of parameters of the dynamics $(\mathbf{v}, \mathbf{\Sigma}, \tau_{\alpha}, \tau_d)$ on the accuracy of the reconstruction.
\subsubsection{The estimator $\hat \tau_\alpha$ performs well, in the case of realistic 2D-TIRF, where $\frac{l}{l_u}\approx0.42$}
The estimator $\hat \tau_\alpha$ is proposed in Eq. \ref{eq:defHatTauAlpha}. Here we test how it performs with different spatio-temporal sampling $(l/l_u, T_S)$, and different birth rate $\lambda$ and death rate $\tau_d$.

By its definition in section \ref{sec:likelihood}, $\tau_\alpha$,  the rate of "spontaneous input" induced by particles born in the unobserved region and reach the border of the ROO, is not a preset parameter. A reference of the "true" value of $\tau_\alpha$ is given by $\frac{N_{l_u}}{T_s}$, where $N_{l_u}$ denotes the number of tracklets born in the region $]-l_u,0]\times[0,H]$ and reached the border $\{0\}\times[0,H]$, $l_u$ is the width of the unobserved region.

\begin{figure}[t!]
\centering
\hspace{-0.5cm}\includegraphics[scale=0.7]{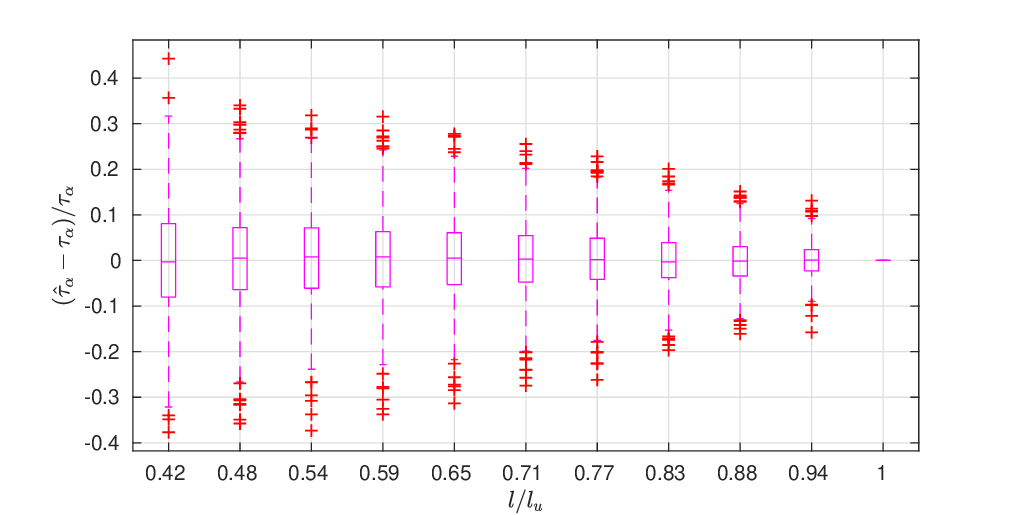}
\caption{At $l/l_u=0.42$, which corresponds to the realistic situation in 2D-TIRF, more than half of the trials presents a relative error smaller than $10\%$. The proposed estimator $\hat{\tau}_\alpha$ is unbiased and the variance decreases as $l/l_u$ increases. }
\label{fig:tau_alpha_portion}
\end{figure}

Next, we test the robustness of the estimator $\hat \tau_\alpha$ w.r.t. $l/l_u$ (Fig. \ref{fig:tau_alpha_portion}). To avoid the influence of $T_S$ on the consistency of the estimator, $T_S$ is set to be long enough as 30 min. We can conclude that, naturally, the more the observed area is larger, better is the performance of the estimator $\hat \tau_\alpha$. In the case of the simulation of the real situation, where $l/l_u=0.42$, the estimator works reasonably good. 

\begin{figure}[t!]
\centering
\hspace{-0.5cm}\includegraphics[scale=0.7]{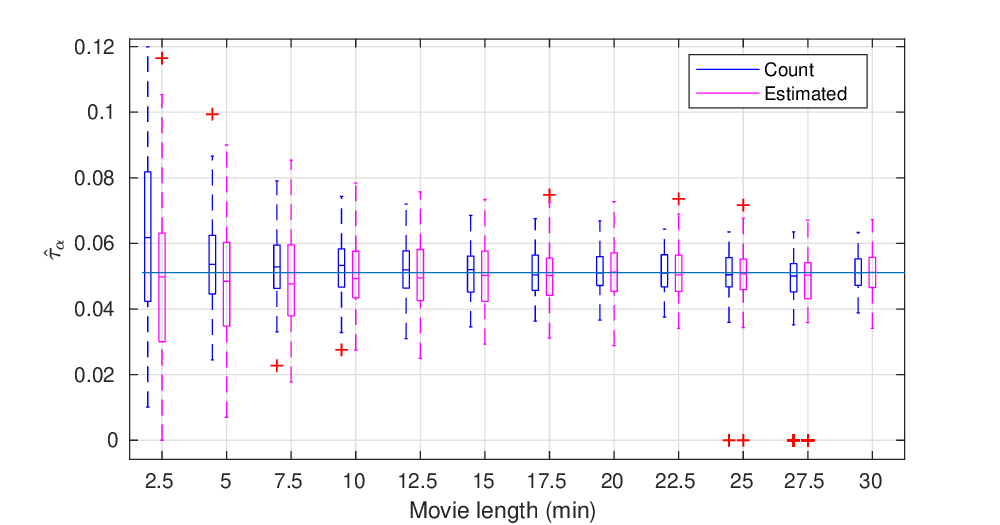}
\caption{The counted 'ground truth' $\tau_\alpha$ and the estimated $\hat{\tau}_{\alpha}$ obtained by movies of different duration, varying from 2.5 min to 30 min. Blue boxes (resp. Magenta boxes) correspond to the 'ground truth', i.e. the counted value (resp. the estimator $\hat{\tau}_\alpha$). The blue horizontal line represents the 'ground truth' value when $T_S=30$ min.}
\label{fig:tau_alpha_time}
\end{figure}

Following, we test the robustness of $\hat{\tau}_{\alpha}$ w.r.t. $T_S$ (Fig. \ref{fig:tau_alpha_time}). This test is essential because in reality it is impossible to use a 30-min movie, because of technical issues like photobleaching of fluorophores and natual growth in living samples. At this stage, the propotion of observed and unobserved region $l/l_u$ is set to $0.42$. $T_S$ varies from 2.5 min to 30 min. In Fig. \ref{fig:tau_alpha_time}, it can be noticed that the reference 'ground truth' of $\tau_\alpha$ (blue boxes) decreases as $T_S$ lengthens. Actually, the reference is only a pseudo 'ground truth'. It is sensible to $T_S$  when $T_S$ is small and it converges as $T_S \rightarrow \infty$. The distributions of counted 'ground truth' and estimator become close to each other for $T_S \geq 10$ min. 

\begin{figure}[t]
\centering
\hspace{-0.5cm}\includegraphics[scale=0.76]{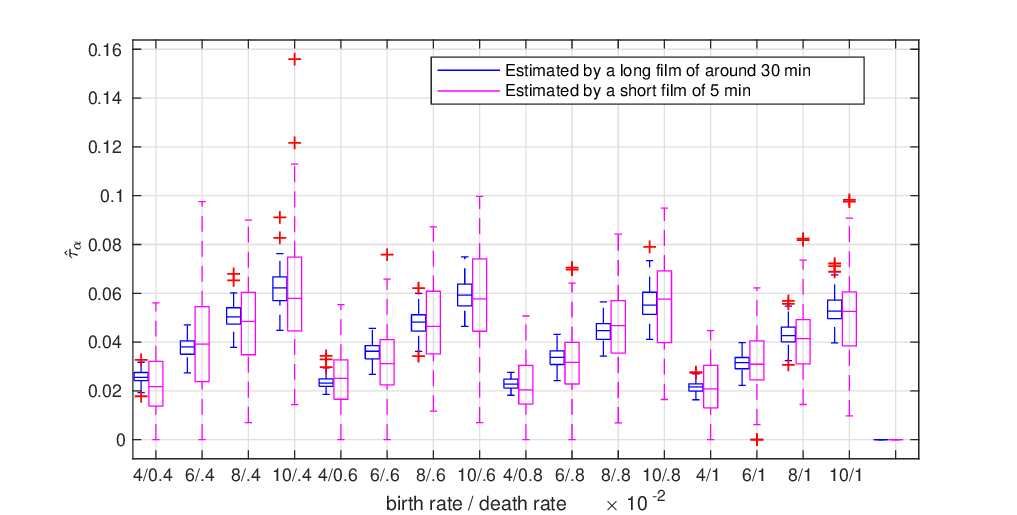}
\caption{The estimation of arrival rate $\tau_\alpha$ w.r.t. different $\lambda$ and $\tau_d$. For example, when $\lambda=0.04, \tau_d = 0.004,$ the median value of $\hat{\tau}_\alpha$ is around 0.025, which means that at each moment, the probability that a particle born in the invisible zone arrives at $\{-l\}\times[0,H]$ is around 0.025.}
\label{fig:hat_tau_alpha_5min_delta_t0_25}
\end{figure}

The absolute value of $\tau_\alpha$ depends on $\lambda$ and $\tau_d$. Fig. \ref{fig:hat_tau_alpha_5min_delta_t0_25} displays for different combinations of $\lambda$ and $\tau_d$, the estimations of $\hat{\tau}_\alpha$ by 5-min movies (magenta) and 30-min movies (blue). It shows that $\tau_\alpha$ increases linearly as the birth rate $\lambda$ increases, and decreases slightly linearly as the death rate $\tau_d$ increases.
\subsubsection{The estimator $\hat{\tau}_d$ is unbiased and performs reasonably well with 5-min movies}
\begin{figure}[t]
\centering
\hspace{-0.5cm}\includegraphics[scale=0.76]{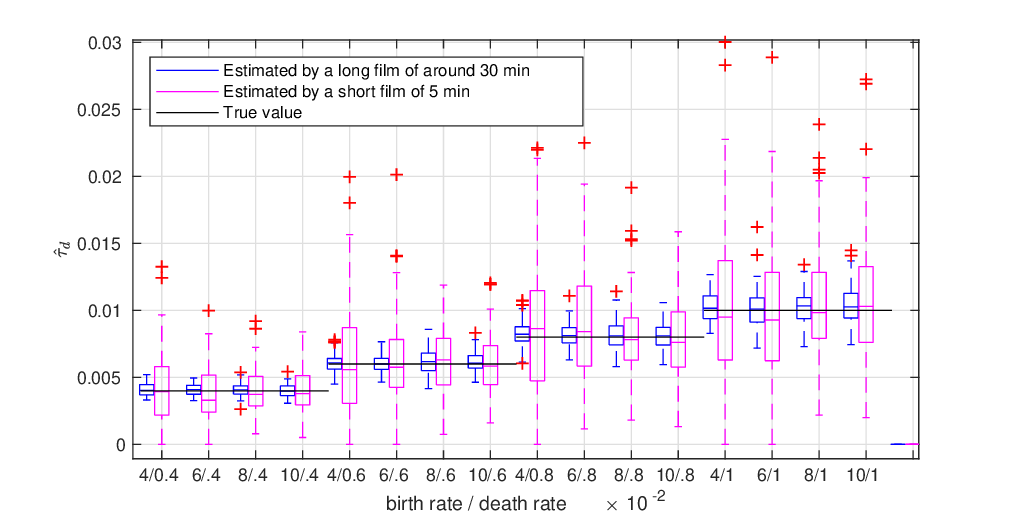}
\caption{$\hat{\tau}_d$ with different $\lambda$ and $\tau_d$. The estimator is unbiased. The variance of the estimator is larger with shorter movies (magenta). For a given $\tau_d$, when birth rate $\lambda$ increases (e.g. the first four boxes), then the number of particles also increases, in consequence, the variance decreases.}
\label{fig:hat_tau_d_5min_delta_t0_25}
\end{figure}

As explained in section \ref{sec:estimators}, $\hat{\tau}_d$ is a rather classical estimator. Fig. \ref{fig:hat_tau_d_5min_delta_t0_25} shows the estimator with 5-min movies (magenta) and 30-min movies (blue) respectively. It confirms that the estimator is unbiased. Black horizontal lines represent the true value of $\tau_d$. Naturally, the variance is bigger with shorter movies.

\subsubsection{The choice of $T_S$}
According to Figs. \ref{fig:hat_tau_alpha_5min_delta_t0_25} and \ref{fig:hat_tau_d_5min_delta_t0_25}, when $T_S = 30$ min, the estimators of $\tau_\alpha$ and $\tau_d$ perform well, being converged with small variance. As 30-min movie acquisition is almost infeasible under the situation of fluorescence microscopy, we need to find a compromise with smaller $T_S$ and reasonably good estimators. We tested especially $T_S$=2.5 min and $T_S$=5 min. Comparing the estimation results with 2.5-min movies, we found that $T_S = 5$ min is a good choice to limit the estimation error of $\tau_d$ and to ensure a good connection performance (more details about the experiments for the choice of $T_S$ in Supplementary Materials \ref{appendix_1}).
\subsubsection{The connection procedure works well, even when true parameters are unknown}
\begin{figure}[t]
\centering
\hspace{-0.5cm}\includegraphics[scale=0.76]{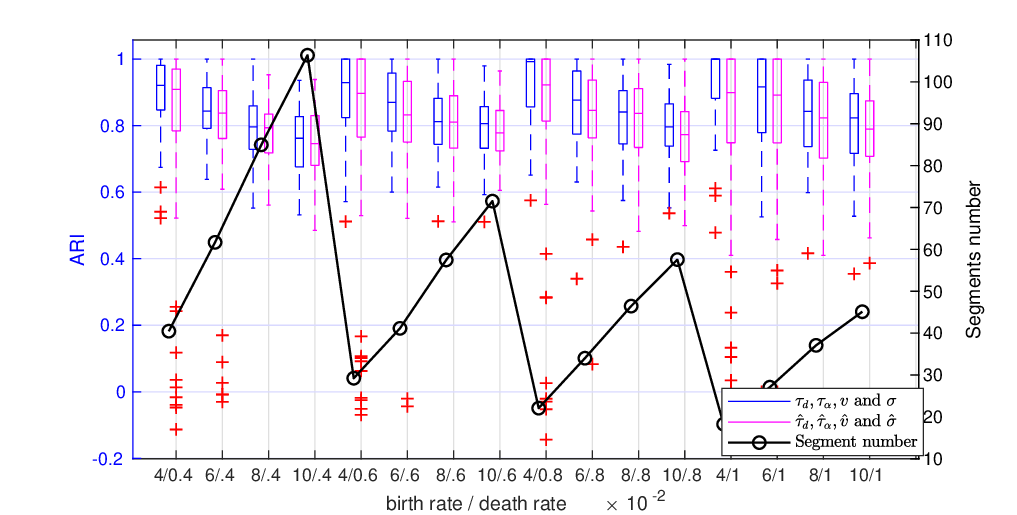}
\caption{Connection performance comparison for different $\lambda$ and $\tau_d$, when $T_S = 5$ min. Blue (resp. magenta) boxes represents ARI values obtained with true parameters $\tau_d, \tau_\alpha, v$ and $\sigma$ (resp. with estimators $\hat{\tau}_d, \hat{\tau}_\alpha, \hat{v}$ and $\hat{\sigma}$). The black line represents the mean number of tracklets in a movie.}
\label{fig:Comp_Ari_5min_delta_t0_25}
\end{figure}

In this part, we assess the performance of the connection algorithm with different parameters $\lambda$ and $\tau_d$. We evaluate as well the impact of the error of the estimator, by using in the connection procedure respectively true parameters $\tau_\alpha, \tau_d, v, \sigma$ and their estimators  $\hat{\tau}_\alpha, \hat{\tau}_d, \hat{v}, \hat{\sigma}$. The duration of movies $T_S$ is set to 5 min. The connection results measured by ARI are presented in Fig. \ref{fig:Comp_Ari_5min_delta_t0_25}. 

Each pair of blue and magenta box represents the connection result of a setting of $\lambda$ and $\tau_d$. The black line represents the mean value of the number of tracklets fluctuating with different settings of $\lambda$ and $\tau_d$. The performance of connection is affected by the number of tracklets in each movie to be connected. The higher the density of tracklets is, the more difficult it is to find the right ones. 

It can be noticed that the ARI value when we use the estimators $\hat{\tau}_d, \hat{\tau}_\alpha, \hat{v}$ and $\hat{\sigma}$, is almost as good as when we use true values for all the parameters. This is an encouraging result as it means that it is feasible to apply the algorithm in real image sequences. When the number of tracklets is around 20 (e.g. $\lambda = 0.04$ and $\tau_d=0.008$), the median values of ARI are higher than 0.9, showing a promising connection performance. Even for the case with the highest particle density, when the average number of tracklets reaches 100 ($\lambda = 0.1$ and $\tau_d=0.004$), the median value of ARI is still higher than 0.7. 

\subsubsection{The connection procedure is robust even when each particle moves at different speed (but with constant speed along a trajectory). However $v$ and $\sigma$ should be well estimated}
In the  previous experiments, all trajectories are generated with the same speed $v$ and standard error $\sigma$. In this section, we design experiments to test the performance of the connection algorithm, when the drift $v$ varies from particle to particle, $v_x\sim \mathcal{U}(0.5,0.9)$. In one movie, as all particles are in the same environment, there is no obvious reason for different particles to have different $\sigma$ . Therefore the standard error $\sigma$ is set to be constant for particles in one movie. However, we test in independent movies, when $\sigma = 0.2, 0.3$ or 0.4, the influence of $\sigma$ on the performance of connection procedure. Other parameters to be specified are the angle between the direction of the motion and the circumferential direction of the cylinder, $\theta = 0.15 (\approx 8.6^{\circ})$, $v_y=\tan(\theta) v_x$, $\sigma_y = \tan(\theta) \sigma_x$, $\sigma_x=\sigma$, birth rate and death rate are fixed, with $\lambda = 0.08$ and $\tau_d=0.02$. 

\begin{figure}[t!]
\centering
\hspace{-0.5cm}\includegraphics[scale=0.7]{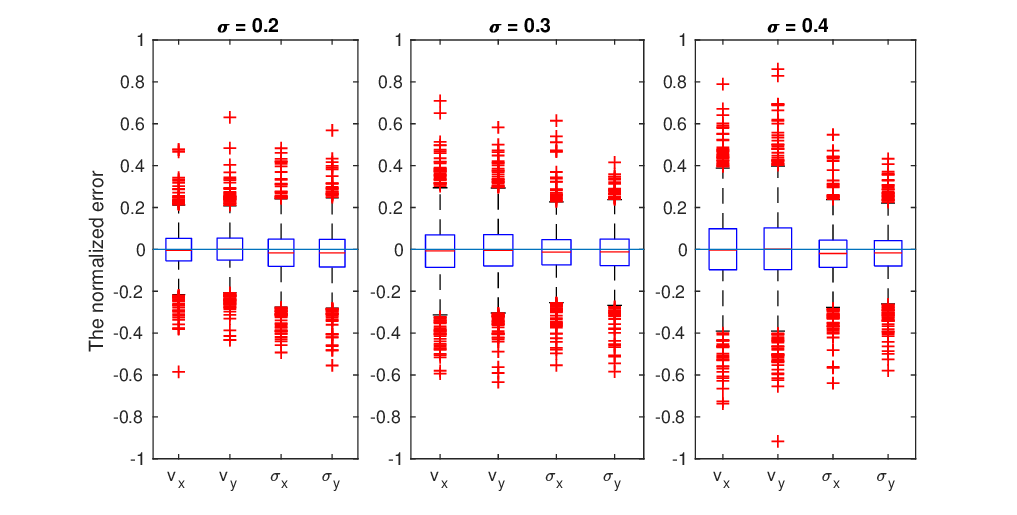}
\caption{The NEs of $v_x, v_y, \sigma_x$ and $\sigma_y$ in cases where $\sigma = 0.2, 0.3$ and $0.4$.}
\label{fig:estim_v_sigma1}
\end{figure}

\begin{figure}[t!]
\centering
\hspace{-0.5cm}\includegraphics[scale=0.7]{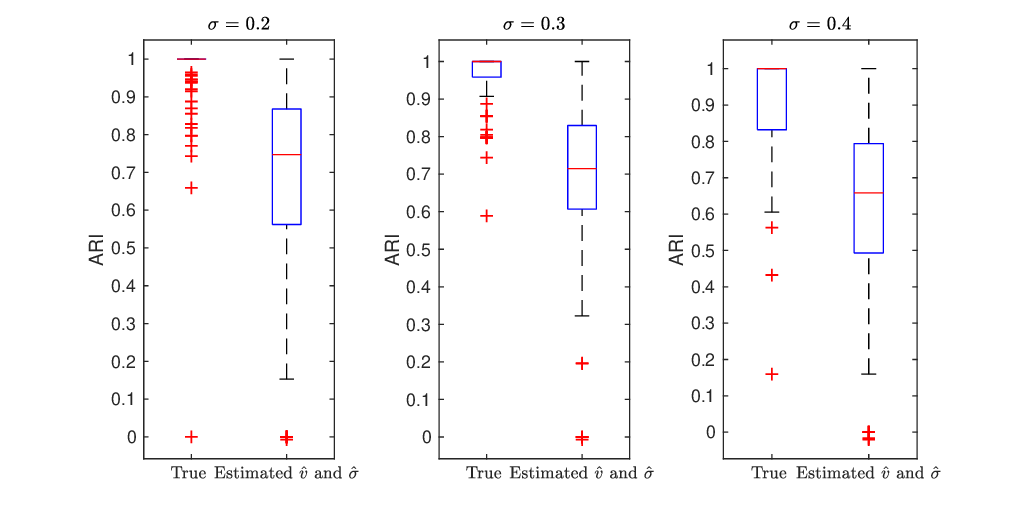}
\caption{Comparison of connection performance measured by ARI, when the procedure takes true $v$ and $\sigma$ or estimated $\hat{v}$ and $\hat{\sigma}$, in three experiments where $\sigma = 0.2, 0.3$ and $0.4$ respectively.}
\label{fig:Ari_5min_delta_t0_25_v_sigma}
\end{figure}

The normalized error (NE) of an estimator is defined by the error of the estimator normalized by its ground truth. For example, the NE of $v_x$ equals to $\frac{v_x-\hat{v}_x}{v_x}$. In Fig. \ref{fig:estim_v_sigma1}, the NEs of $\hat{v}_x, \hat{v}_y, \hat{\sigma}_x$ and $\hat{\sigma}_y$ when $\sigma$ takes different values are presented. It shows that when $\sigma$ increases, the variance of $\hat{v}_x$ and $\hat{v}_y$ increases.

For tracklets connection, we compare the results when true values of $v$ and $\sigma$ or when the estimated value $\hat{v}$ and $\hat{\sigma}$ respectively are taken by the connection procedure. The experiments are carried under three situations, when $\sigma = 0.2, 0.3,$ and 0.4. The results in Fig. \ref{fig:Ari_5min_delta_t0_25_v_sigma} shows that whether using true $v$ and $\sigma$ or estimated value $\hat{v}$ and $\hat{\sigma}$, the performance measured by ARI degrades when $\sigma$ increases. When the standard error $\sigma=0.2$, using true $v$ and $\sigma$, the median value of ARI reaches to 1. When using the estimated $\hat{v}$ and $\hat{\sigma}$, the median value of ARI is approximately 0.75. It can be concluded that the estimation of $v$ and $\sigma$ has an impact on the performance of the algorithm.

\subsection{Analysis of the connection results}
\subsubsection{An example of tracklets connection}
\begin{figure}[t]
\hspace{-1cm}\includegraphics[scale=0.38]{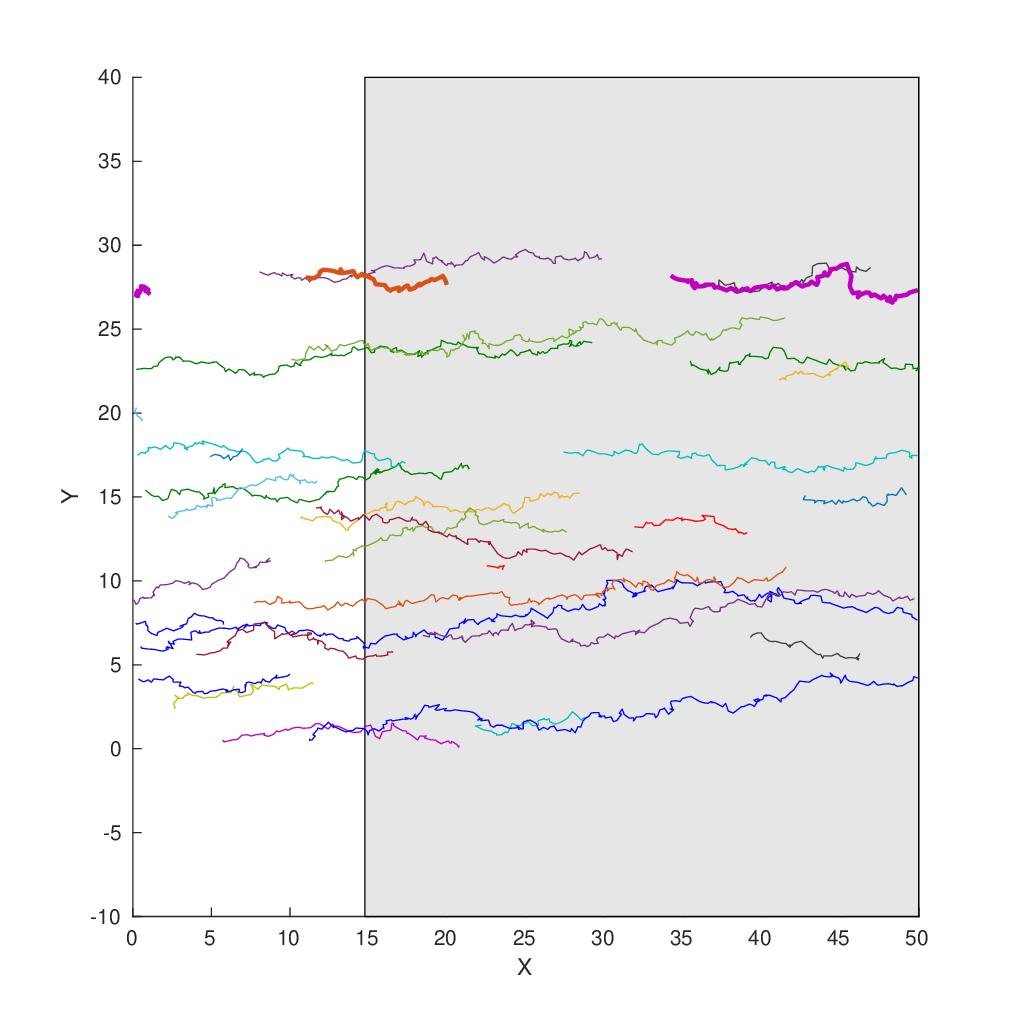}
\hfill \includegraphics[scale=0.38]{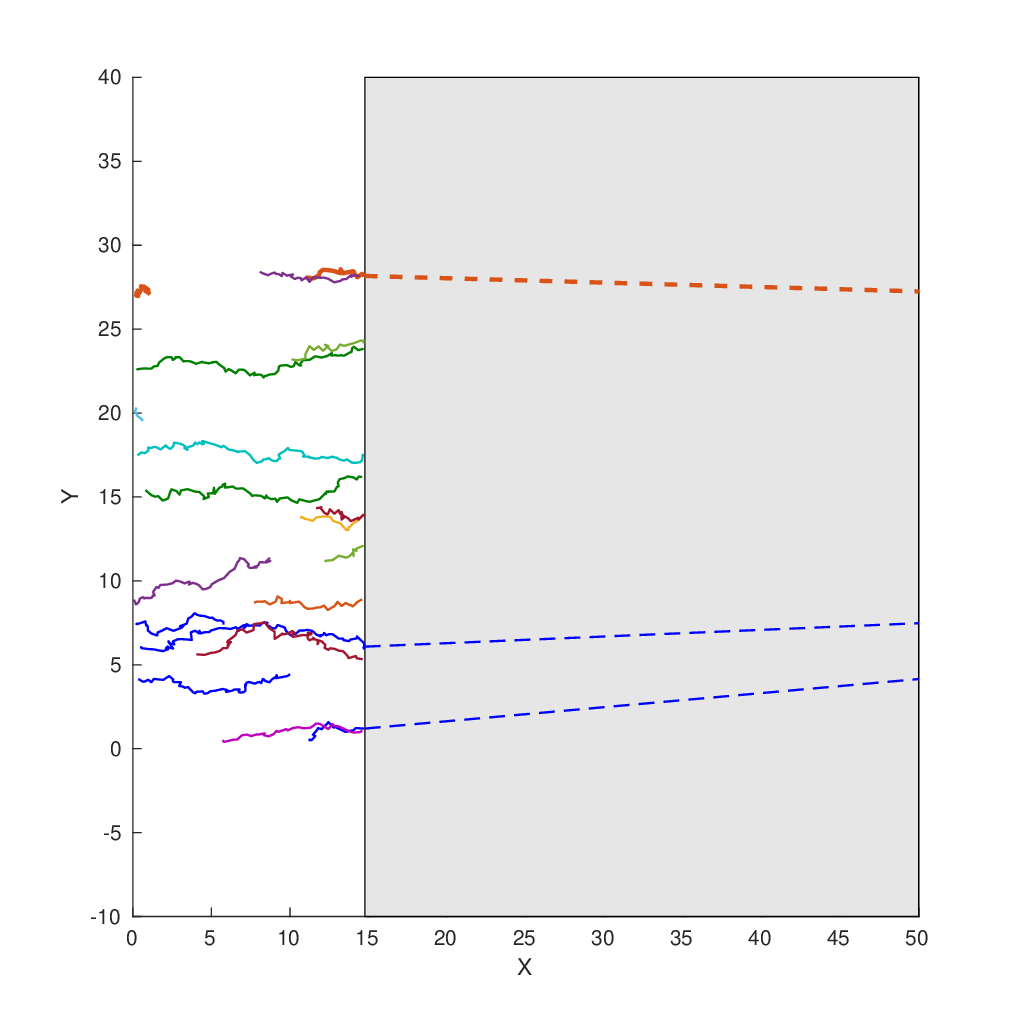}
\caption{Left: Trajectories in one movie; Right: the connection results.}
\label{tracks_conn}
\end{figure}

Figure \ref{tracks_conn} shows, on the left, trajectories in a movie and on the right, the results of tracklets connection. The path from an output to the matched input is represented by the dashed straight line, as we don't know how exactly the particle went through the hidden zone. The only wrong connection corresponds to the bold line. Compared with the figure on the left, we can find the realization of these two tracklets. In reality, the orange bold tracklet disappeared at the hidden region and the bold purple tracklet appeared nearby and entered into the observed zone. 

\begin{figure}[t]
\centering
\hspace{-0.5cm}\includegraphics[scale=0.65]{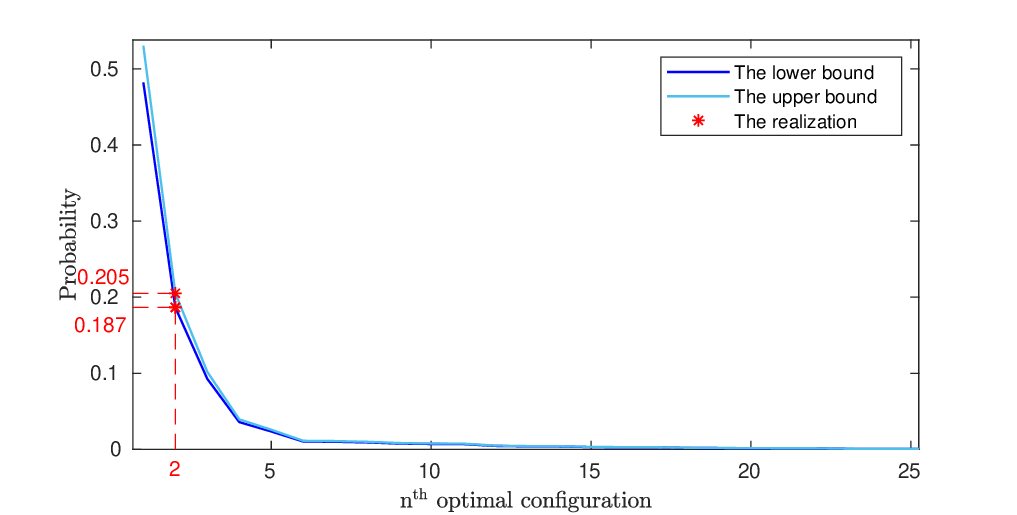}
\caption{The probability of $n$\textsuperscript{th} optimal configuration and the probability of the realization.}
\label{prob_neme}
\end{figure}

In fact, not only the optimal configuration can be calculated, but also the most likely alternative configurations in decreasing order of probability (Fig. \ref{prob_neme}). It should be noticed that the optimization algorithm tries to minimize $K(c)=-\log{Q(c)}$, instead of finding the $c^*$ maximizing $P(c)$. It costs too much to obtain the probability $P(c^*)$, as it requires the enumeration of all the possible configurations $c\in\mathcal{C}$ (Eq. \ref{eq:LimEpsQ}). However, The number of configurations can be determined in order to guarantee that the sum of the probability of these configurations will be greater than a given threshold (see Supplementary Materials \ref{appendix_2}). As a result, we can obtain lower and upper bounds for the probability.

For this example, we see that the second most likely configuration corresponds to the realization of trajectories, $0.187<P(c^t) < 0.205$ according to the algorithm (Fig. \ref{prob_neme}). Combining with Fig. \ref{tracks_conn}, the optimal configuration found by the algorithm, committing one connection error, does not correspond to the realization. In section \ref{sm:Analyse_ari}, we evaluated the connection error caused by randomness.

%
%

\subsubsection{The number of rotations around the cylinder}
Once the connection procedure is achieved, we can address the question of the number of rotations of a particle around the cylinder. In the context of simulation, the death rate $\tau_d$ and the dynamic velocity $v_x$ are known. Accordingly, the value of the number of rotations is known to be equal to $\frac{v_xT_d}{L},$ where $T_d \sim \mathcal{E}(\tau_d)$ ensures a theoretical expectation value of $\frac{v_x}{\tau_d L}.$ 
By counting the tracklets for each trajectory, we can obtain a proxy of the number of rotations around the cylinder.

\begin{figure}[ht]
\hspace{-1cm}\includegraphics[scale=0.8]{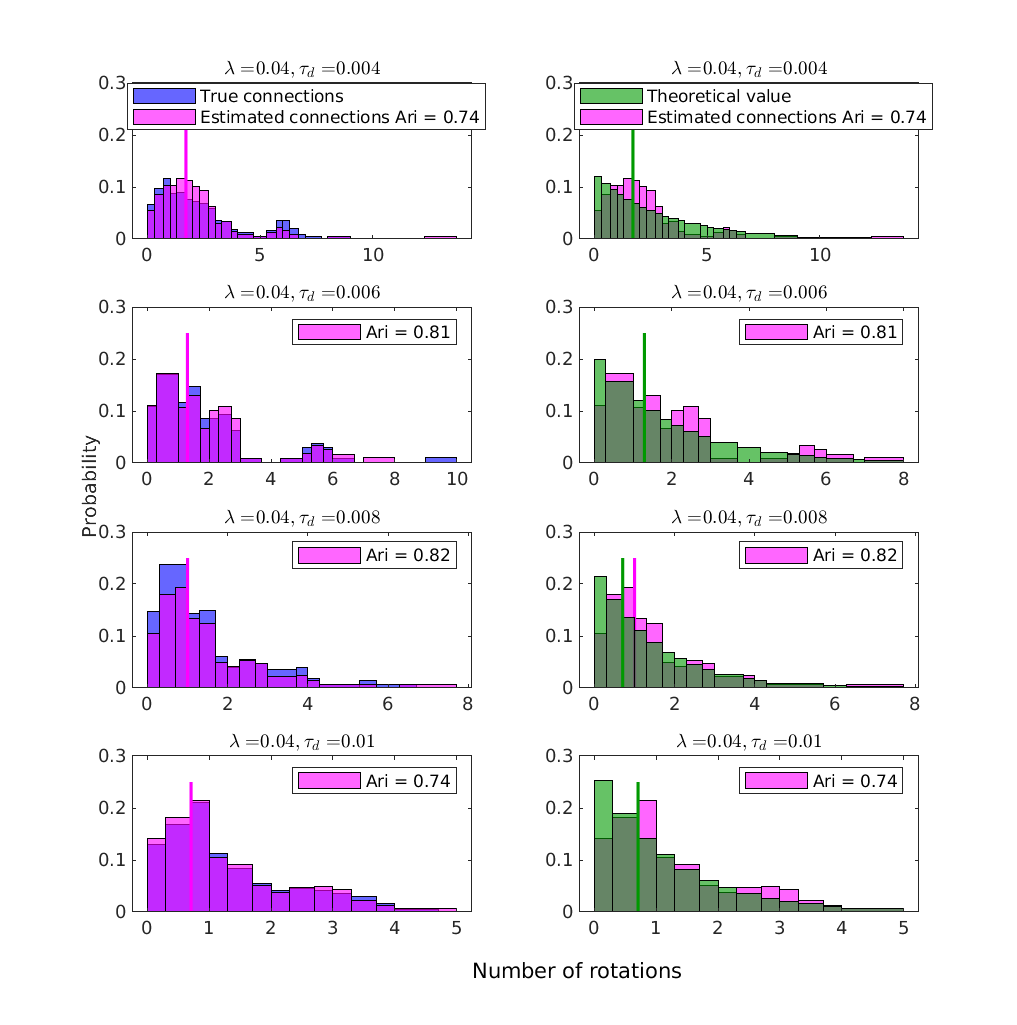}
\caption{The density distribution of 'number of rotations'. Each row represents the result for a different $\tau_d$ value.}
\label{Nb_tours}
\end{figure}

In Fig. \ref{Nb_tours}, $\lambda$ is set to 0.04 and different values of $\tau_d$ between $0.004$ and $0.01$ are evaluated. Blue bars represent the distribution of the number of rotations of true connections. The magenta bars to display the distribution of the number of rotations estimated by the connection procedure. The corresponding ARI values, indicating the connection accuracy, are given as well. The density of the theoretical values of the number of rotations is presented in green color. The vertical lines represent the median values of the corresponding distribution. Overall, when $\tau_d$ is small, the median value of number of rotations is higher and the distribution has a heavier tail. In general, the distributions with all three colors are similar to each other.

\section*{Conclusion}

In this paper, we proposed a probabilistic framework and a computational approach  with no hidden parameter to connect tracklets from 2D partial observations. We provided several consistent estimators of parameters to automatically drive the connection procedure.
The performance of our procedure is satisfying if we consider the ARI criterion. Moreover, an ordered set of the best reconstructions could also be proposed. The robustness of the procedure has been tested for different drifts, diffusion of the dynamics, and trajectory densities.
Our computational approach can be extended to the case when the drift/speed is not the same for all particles but remains constant along time. In that case, it is straightforward to estimate and classify the drifts before applying our connection procedure to each class of drift since the tracklets with different speeds are not likely to be connected.

After recovering the whole trajectory on the surface of the cylinder, we can have a better understanding of the average duration of a particle, and more accurate statistics about the spatio-temporal organization of particles. The simulation study can also serve as a guideline for the design of experiments.

The connection procedure is tested with a real TIRFM dataset. The experimental results are illustrated in Appendix \ref{appendix_3}. For future works, we plan to investigate more on real TIRFM datasets. Experiments on real data show that the observed region corresponds approximately to one-third of the total surface, which is rather small. However, we have shown that we are able to cope with the hidden region of such size. Nevertheless, several assumptions and approximations need to be further investigated. For instance, 
we assumed spatial homogeneity, suggesting that the particles are born or die uniformly on the membrane surface. 
Moreover, we assumed a memoryless lifetime while dependency with respect to particle ``age'' could be more realistic.


\appendix

\section{An illustration of the connection algorithm applied to real MreB dynamics}\label{appendix_3}
Data obtained using TIRF microscopy of MreB aggregates in \textit{Bacillus subtilis} (\cite{billaudeau2017contrasting}) are considered. A typical movie from this dataset shows several MreB aggregates moving inside one or several cells (see Fig. \ref{fig:illus_partialobservation} and Supplementary Materials 3). The pixel size, frame rate and duration are respectively $\Delta x= \Delta y = 64nm$, $\Delta t = 1s$, $T=2mn$.  Hereafter, we selected one cell to illustrate the application of our algorithm. First, tracklets exhibiting directed motion should be extracted from the movie data, then tracklets should be projected back on the cylinder shape of the cell and unwrapped, eventually the connection algorithm is applied and a list of likelihood decreasing ordered configurations of trajectories connections is presented to the user.

\subsection{Construction of the local cell referential}
\begin{figure}[ht]
\centering
\hspace{-0.5cm}\includegraphics[scale=0.6]{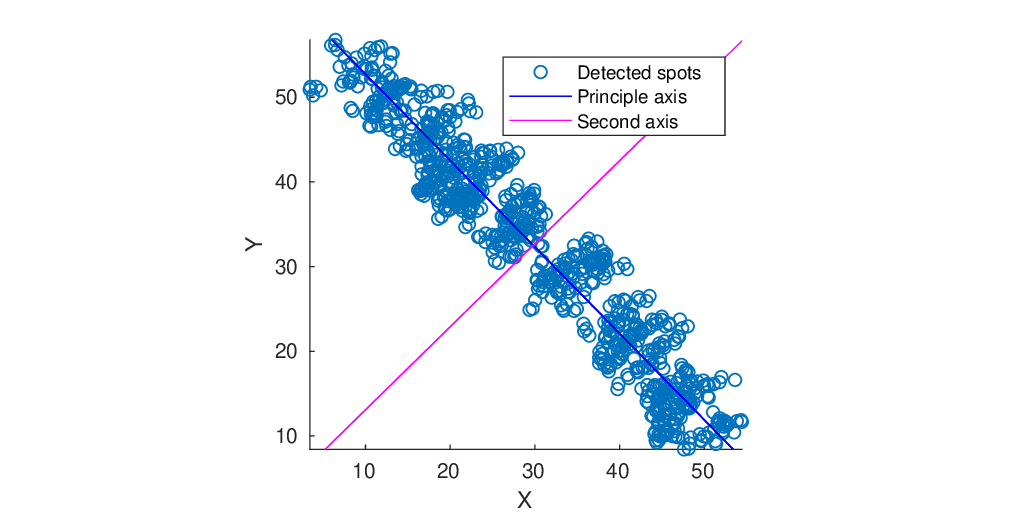}
\caption{The estimation of the local $x-y$ referential for a cell. }
\label{fig:PCA}
\end{figure}

Once MreB aggregates pixels are separated from the background inside each image of the movie,  a bounding box is drawn around a given cell and a local $\textbf{x-y}$ referential is estimated using Principal Component Analysis (PCA) on the coordinates of pixels belonging to aggregates (Fig. \ref{fig:PCA}). The $\textbf{z}$ coordinate of an aggregate is inferred using as a prior the cylinder shape of the bacteria and its radius, $R$, so $z(x,y)=R-\sqrt{R^2-x^2}$.

\subsection{Tracking and selection of aggregates in the observed region}
\begin{figure}[ht]
\hspace{-0.5cm}\includegraphics[scale=0.76]{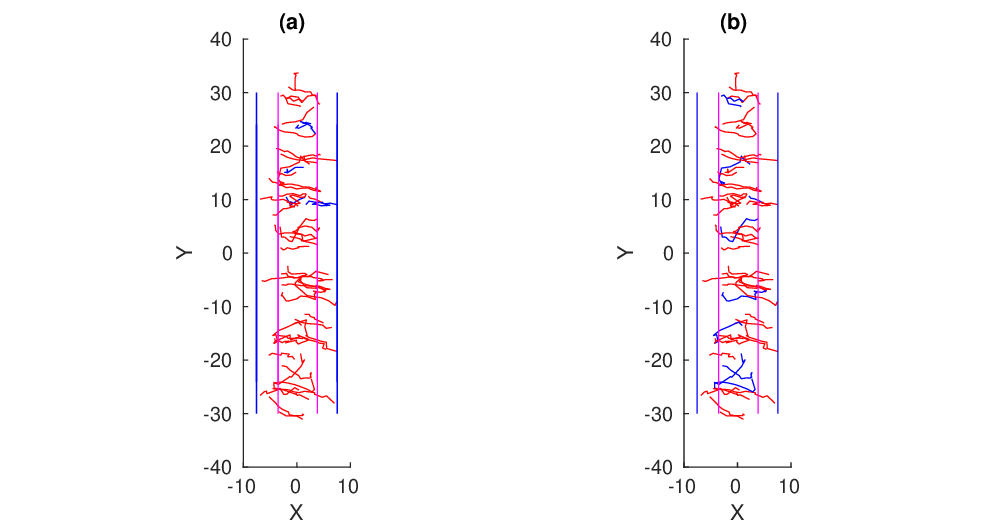}
\caption{The tracklets classification. (a) MSD classification. (b) STP classification. Brownian tracklets (blue), Directed tracklets (red). Blue lines represent the border of the visible region. Magenta lines represent the 0.1 quantile and the 0.9 quantile of \textit{x} coordinate values.}
\label{fig:classifExpe}
\end{figure}

Using U-Track \cite{jaqaman2008robust}, MreB aggregates are tracked and constitute a set of tracklets. The automatic  classification of these tracklets in three classes, respectively Brownian, subdiffusive and, directed motion is done using two algorithms: the classical MSD algorithm and a recent algorithm  (\cite{briane2018statistical}). The tracklets classified as directed motion by either one of the two algorithms are selected for the application of our connection algorithm (Fig. \ref{fig:classifExpe}). The tracklets were projected  back on the cylinder and unwrapped, as explained in the technical part of the paper.  As we can see, only a few aggregates crossed the borders of the visible region. Others aggregates, according to our definitions are born or die in the visible region, which is not true. When an aggregate approaches the borders, its intensity becomes weak as it is farther from the support plane, and less excitation light penetrates higher z-position in TIRF microscopy settings. As a result, the detection algorithm fails to detect the aggregates when they approach the borders.

\subsection{The connection of tracklets}
All the selected tracklets crossing the magenta lines in Fig. \ref{fig:classifExpe} are considered.

\begin{figure}[ht]
  \centering
\begin{tabular}{cc}
\hspace{-0.5cm}\includegraphics[scale=0.36]{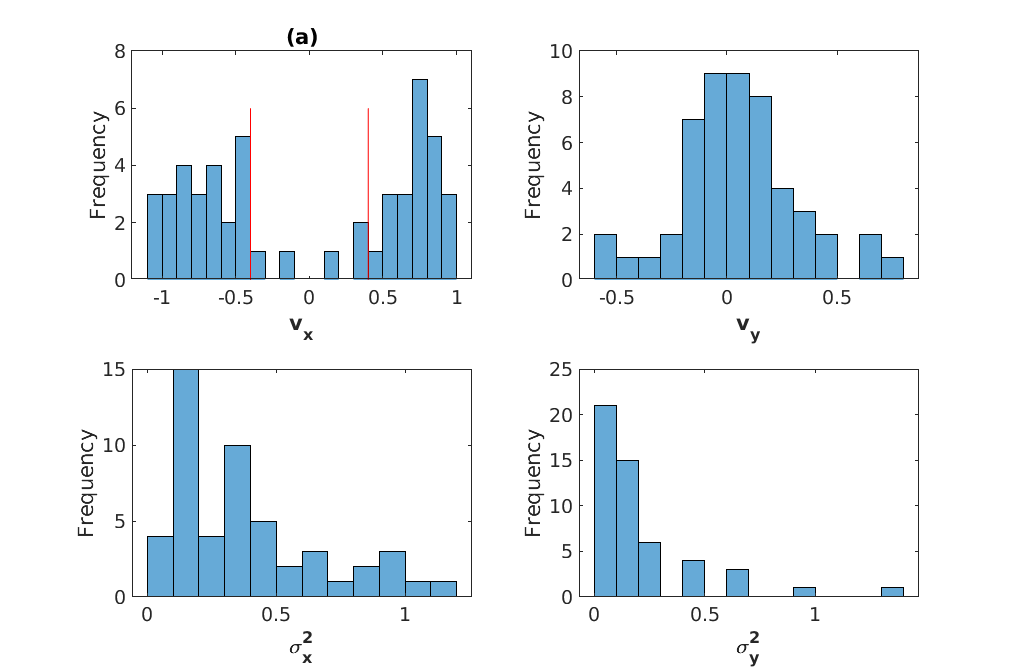} &\includegraphics[scale=0.36]{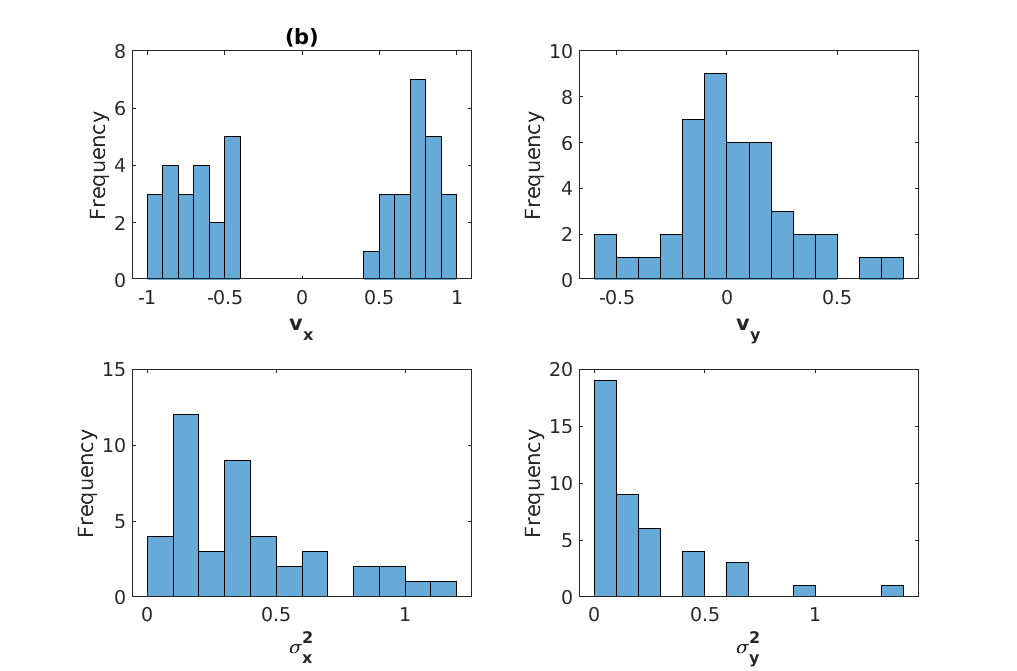}
\end{tabular}
\caption{The distribution of drift and variance in the selected tracklets population. (a) without filtering. (b) after filtering. }
\label{fig:SpeedExpe}
\end{figure}

First, the speed and diffusion are estimated (Fig. \ref{fig:SpeedExpe}) for each tracklet, respectively. Two populations of tracklets evolving in opposite $x$ directions are identified. These two populations are considered one after the other in the connection procedure. Tracklets corresponding to speed lower than 0.4 are filtered out.

For the population of tracklets associated with positive (resp. negative) $v_x$, death rate $\tau_d$ is estimated as 0.0691 (resp. 0.0756). The arrival rate $\tau_{\alpha}$ is estimated as 0.0310 (resp. 0.0220).

$\dagger$ \textbf{tracklets of positive speed $v_x$}

\begin{figure}[ht]
  \centering
\begin{tabular}{ccc}
\hspace{-0.5cm}\includegraphics[scale=0.44]{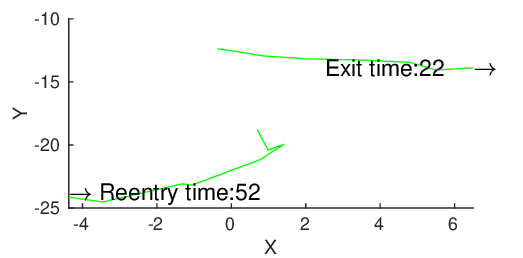} &\includegraphics[scale=0.44]{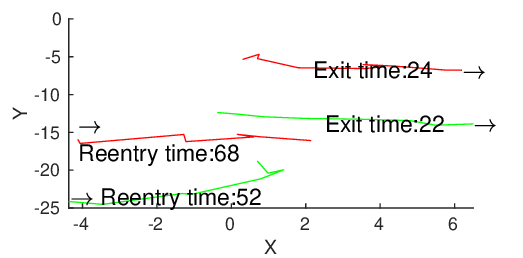}&\includegraphics[scale=0.44]{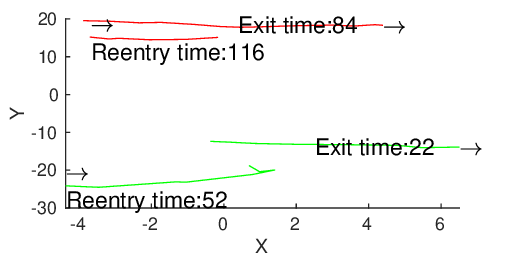}
\end{tabular}
\caption{Some optimal configurations for the population of positive speed $v_x$. From left to right, first, third and sixth better configurations. Connected tracklets are drawn with the same color.}
\label{fig:dataset1}
\end{figure}

The first, fifth, seventh and eighth optimal configuration suggests one connection. The second suggests that there is no connection. The third, fourth and sixth configurations suggest two connections.  Some of these configurations are shown in Fig. \ref{fig:dataset1}.

$\dagger$ \textbf{tracklets of negative speed $v_x$}

\begin{figure}[ht]
  \centering
\hspace{-0.5cm}\includegraphics[scale=0.40]{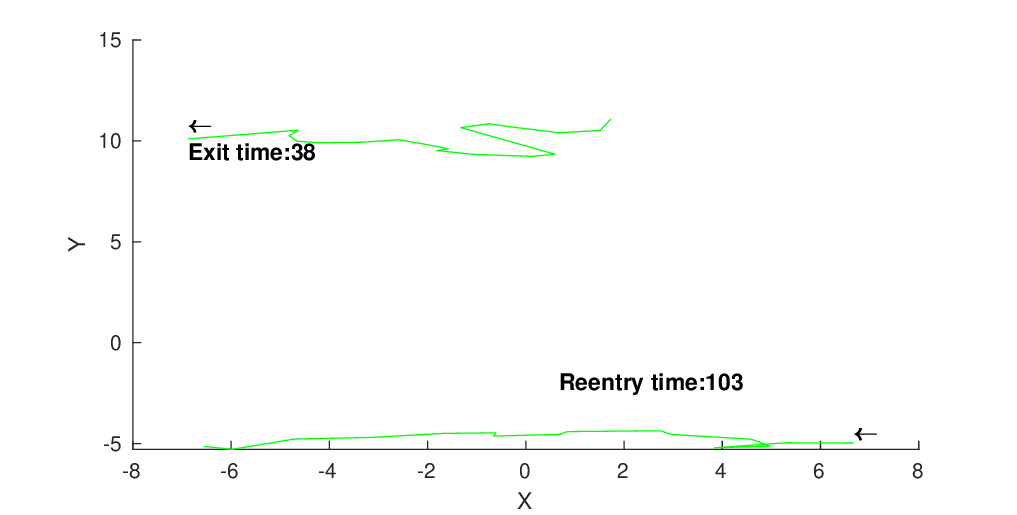}
\caption{Second optimal configuration for the population of negative speed $v_x$.}
\label{fig:dataset2}
\end{figure}

The first optimal configuration suggests no connection. The second one suggests one connection Fig. \ref{fig:dataset2}.

\begin{figure}[ht]
\centering
\hspace{-0.5cm}\includegraphics[scale=0.38]{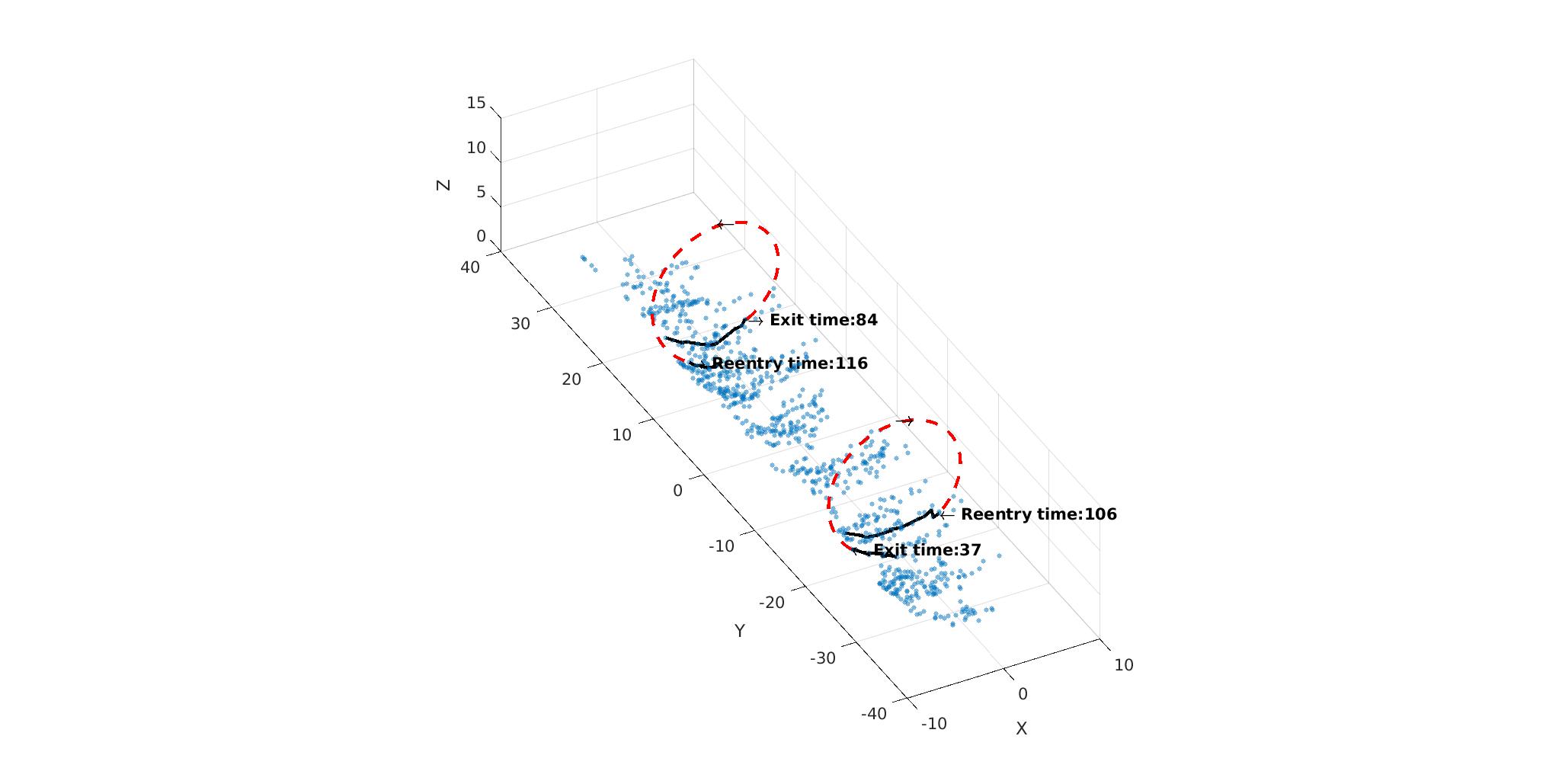}
\caption{The 3D reconstruction of tracks. The centroids of aggregates are represented as blue squares.The arrows indicates the direction of motion. The full blue lines represent tracklets crossing the observed region. The dotted red lines represent the simplified extrapolation of aggregate motion in the invisible region.}
\label{fig:recons_3D_visual}
\end{figure}

In Fig. \ref{fig:recons_3D_visual} we show a 3D reconstruction of the aggregates and two tracks that could correspond to aggregates doing more than one loop around the cylinder surface of the cell. For the positive (resp. negative)  speed set of tracklets, the eighth (resp. second) optimal solution was selected.

\section*{Acknowledgements} The authors thank L. Tournier for fruitful discussions about the optimization procedure, R. Carballido-L\'opez and C. Billaudeau for their inspiring work in MreB studies which triggered this research. This work was partially supported by ANR DALLISH, Programme CES232016.

\bibliographystyle{plainnat}
\bibliography{biblio.bib}

\end{document}


\maketitle

\section{When the length of movies equals to 2.5 min, the estimation errors of $\hat{\tau}_\alpha$ and $\hat{\tau}_d$ cause the failure of the connection procedure}\label{appendix_1}

\subsection{Estimation of birth rate $\tau_\alpha$ and of death rate $\tau_d$}
\begin{figure}[ht]
\centering
\hspace{-0.5cm}\includegraphics[scale=0.76]{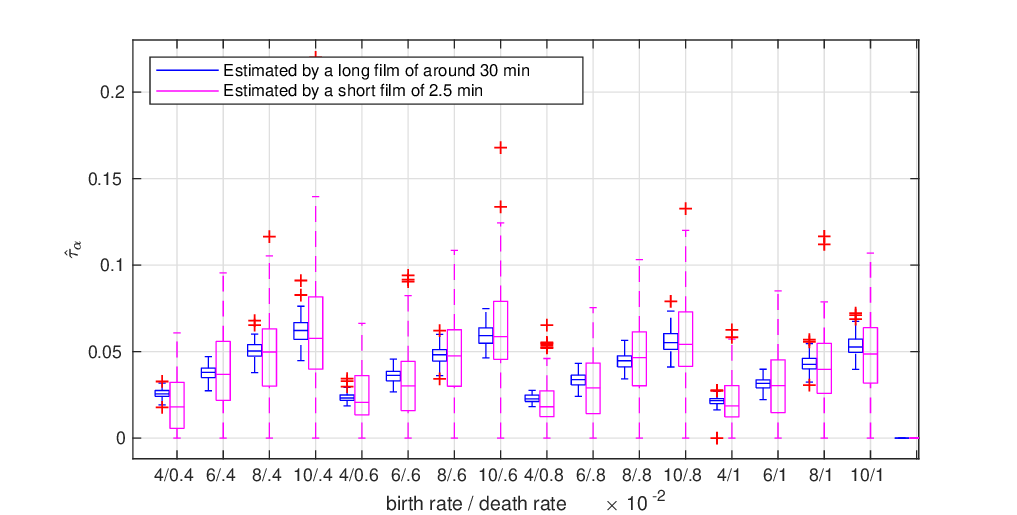}
\caption{The estimation of arrival rate $\tau_\alpha$ with different $\lambda$ and $\tau_d$. Magenta boxes represent $\hat{\tau}_\alpha$ estimated by 2.5-min movies and blue boxes by 30-min movies.}
\label{fig:hat_tau_alpha_2_5min_delta_t0_25}
\end{figure}

\begin{figure}[ht]
\centering
\hspace{-0.5cm}\includegraphics[scale=0.76]{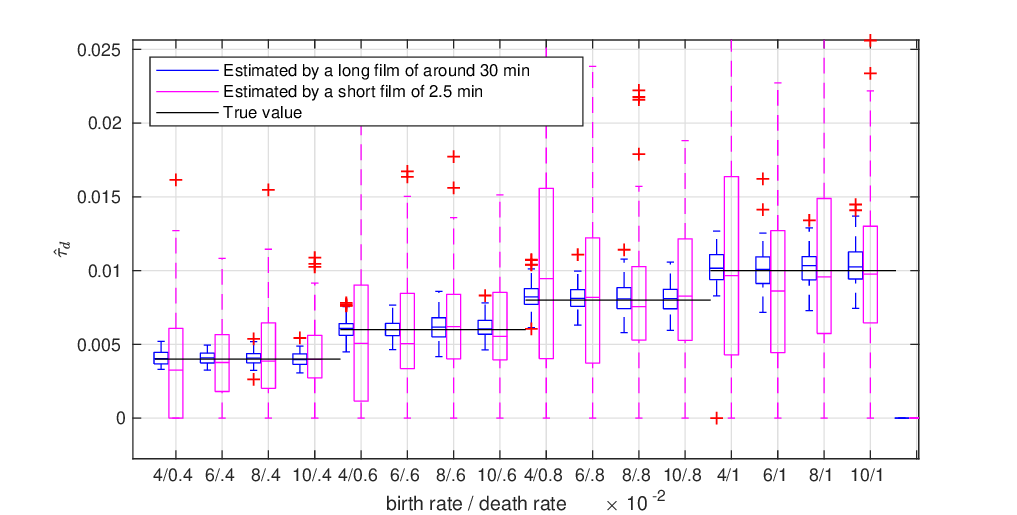}
\caption{The estimation of death rate $\tau_d$ with different $\lambda$ and $\tau_d$. Magenta boxes represent $\hat{\tau}_d$ estimated by 2.5-min movies and blue boxes by 30-min movies. Black horizontal lines represent the true value of $\tau_d$.}
\label{fig:hat_tau_d_2_5min_delta_t0_25}
\end{figure}

The arrival rate $\tau_\alpha$ is estimated from 2.5 min movies and the result is presented in Fig. \ref{fig:hat_tau_alpha_2_5min_delta_t0_25}. Compared to Fig. \ref{fig:hat_tau_alpha_5min_delta_t0_25} where 5-min movies are used, the estimations with 2.5-min movies have bigger variance. With 2.5-min movies, a considerable proportion of the estimations equal to zero. Similar conclusion can be made for $\hat{\tau}_d$ comparing Fig. \ref{fig:hat_tau_d_5min_delta_t0_25} and Fig. \ref{fig:hat_tau_d_2_5min_delta_t0_25}.

It is shown that both in Figs \ref{fig:hat_tau_alpha_2_5min_delta_t0_25} and \ref{fig:hat_tau_d_2_5min_delta_t0_25}, the estimators $\hat{\tau}_\alpha$ and $\hat{\tau}_d$ have many zero values. This phenomenon will deny the birth and death of particles, and force the connection of observed tracklets. The result of connection is presented in the next section and it shows that these biases of parameter estimation have severe influence on the tracklets connection performance. 

\subsection{Evaluation of the connection procedure}
\begin{figure}[ht]
\centering
\hspace{-0.5cm}\includegraphics[scale=0.5]{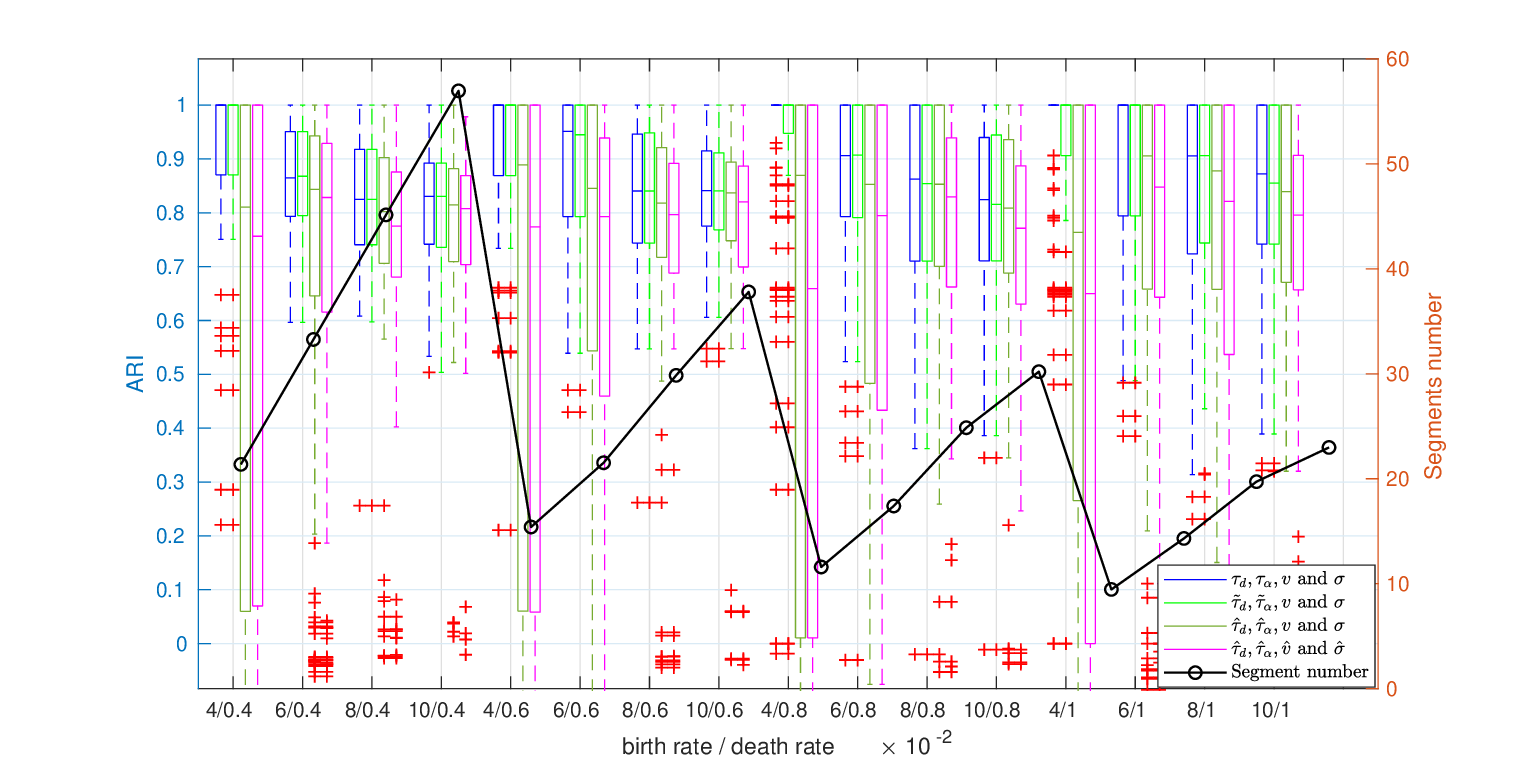}
\caption{Connection performance comparison. Blue (resp. green, dark green and magenta) boxes represent ARI values obtained with parameters ($\tau_d, \tau_\alpha, \mathbf{v}, \sigma$) (resp. ($\tilde{\tau}_d, \tilde{\tau}_\alpha, \mathbf{v}, \sigma$), ($\hat{\tau}_d, \hat{\tau}_\alpha, \mathbf{v}, \sigma$) and ($\hat{\tau}_d, \hat{\tau}_\alpha, \hat{\mathbf{v}}, \hat{\sigma}$)). Light green boxes, representing the result when $\tilde{\tau}_\alpha, \tilde{\tau}_d$ (estimators with 30-min movies) and true value $v, \sigma$ are used, show performance as good as the blue boxes, where true parameters values are used. However, the dark green boxes, where $\hat{\tau}_\alpha, \hat{\tau}_d$ (with 2.5-min movies) and $v, \sigma$ are used, show much degraded results.}
\label{ARI_taux_v_comp1}
\end{figure}

\begin{figure}[ht]
\centering
\hspace{-0.5cm}\includegraphics[scale=0.5]{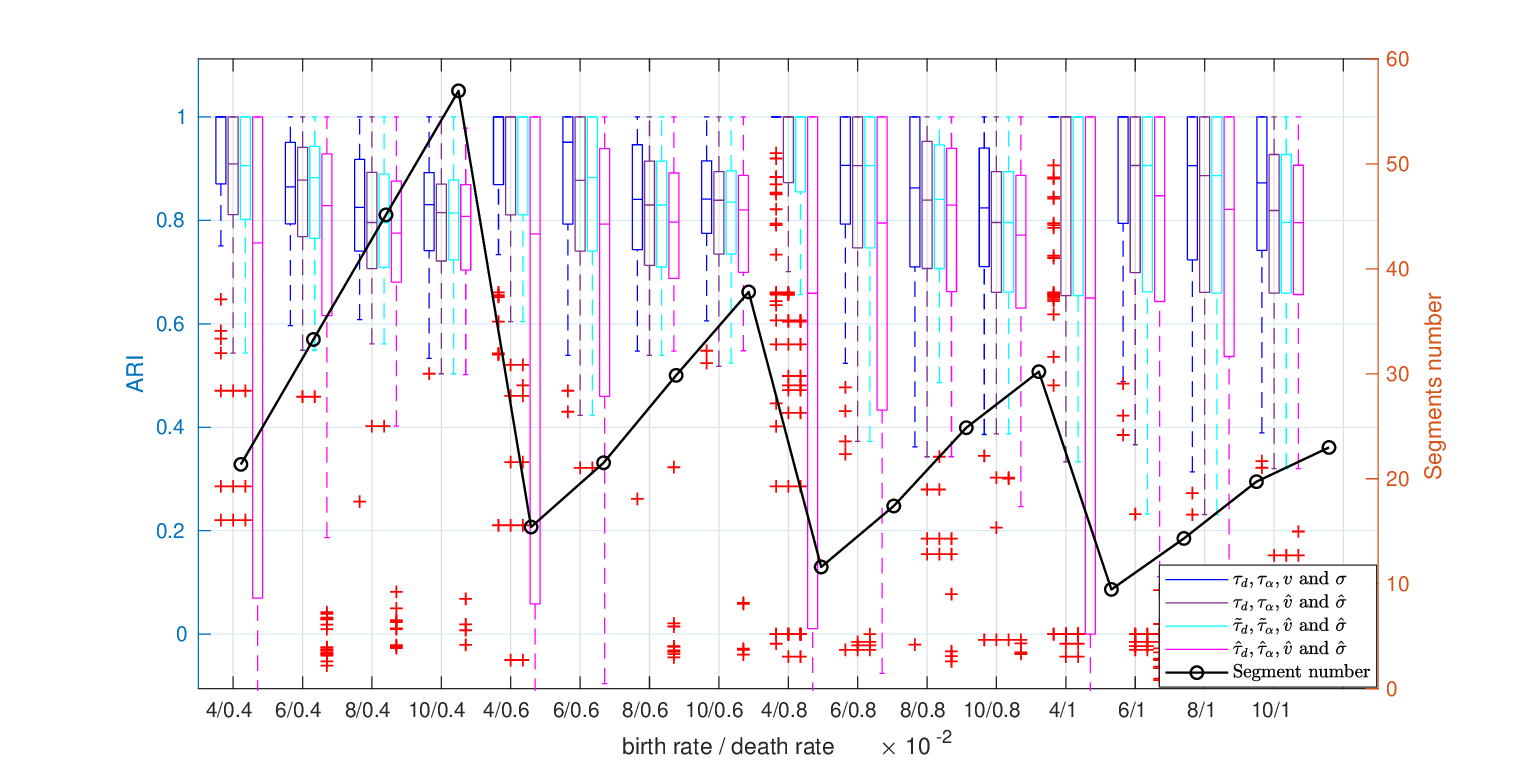}
\caption{Connection performance comparison. Blue (resp. violet, cyan and magenta) boxes represent ARI values obtained with parameters ($\tau_d, \tau_\alpha, \mathbf{v}, \sigma$) (resp. ($\tau_d, \tau_\alpha, \hat{\mathbf{v}}, \hat{\sigma}$), ($\tilde{\tau}_d, \tilde{\tau}_\alpha, \hat{\mathbf{v}}, \hat{\sigma}$), and ($\hat{\tau}_d, \hat{\tau}_\alpha, \hat{\mathbf{v}}, \hat{\sigma}$)). Boxes with blue, violet, and cyan colors are similar, which means that the estimators $\hat{v}, \hat{\sigma}, \tilde{\tau}_\alpha$ and $\tilde{\tau}_d$ do not cause degradation of the connection. Only when $\hat{\tau}_\alpha$ and $\hat{\tau}_d$ are used, shown in magenta boxes, the connection results degrades.}
\label{ARI_taux_v_comp2}
\end{figure}

Figs. \ref{ARI_taux_v_comp1} and \ref{ARI_taux_v_comp2} show the results of tracklets connection measured by ARI, according to different settings of birth rate $\lambda$ and death rate $\tau_d$. To be specified, $\tilde{\tau}_\alpha$ and $\tilde{\tau}_d$ are estimators when $T_S = 30$ min, on the contrary, the estimators obtained with 2.5-min movies are denoted as $\hat{\tau}_\alpha$ and $\hat{\tau}_d$. Concerning the experiment with true parameters (blue boxes in both figures), for all settings of $\lambda$ and $\tau_d$, the connection results are satisfying. However, for the experiments with estimated parameters (magenta boxes in both figures), we see clearly the failure of the connection represented by boxes with very large variance and low ARI values.

To identify which estimators are the main cause of this failure, many intermediate experiments are designed. To remind, $\tilde{\tau}_\alpha$ and $\tilde{\tau}_d$ represent the estimators of $\tau_\alpha$ and $\tau_d$ from 30-min movies, while $\hat{\tau}_d$ and $\hat{\tau}_\alpha$ from 2.5-min movies. Through Fig. \ref{ARI_taux_v_comp1} and \ref{ARI_taux_v_comp2}, it can be concluded that the error of estimators $\hat{\tau}_d, \hat{\tau}_\alpha$ is the main cause of the dramatic decrease of ARI when all the estimators $\hat{\tau}_d, \hat{\tau}_\alpha, \hat{v}$ and $\hat{\sigma}$ are used. 

The accuracy of $\hat{\tau}_\alpha$ and of $\hat{\tau_d}$ increases as the total observed time $T_S$ increases (see Fig. \ref{fig:tau_alpha_time}). Therefore, in the study, we chose movies of 5 min to estimate $\tau_\alpha$ and $\tau_d$ and then to evaluate the connection performance. 

\section{Analysis of errors}\label{sm:Analyse_ari}
\begin{figure}[ht]
\centering
\hspace{-1cm}\hspace{-1cm}\includegraphics[scale=0.43]{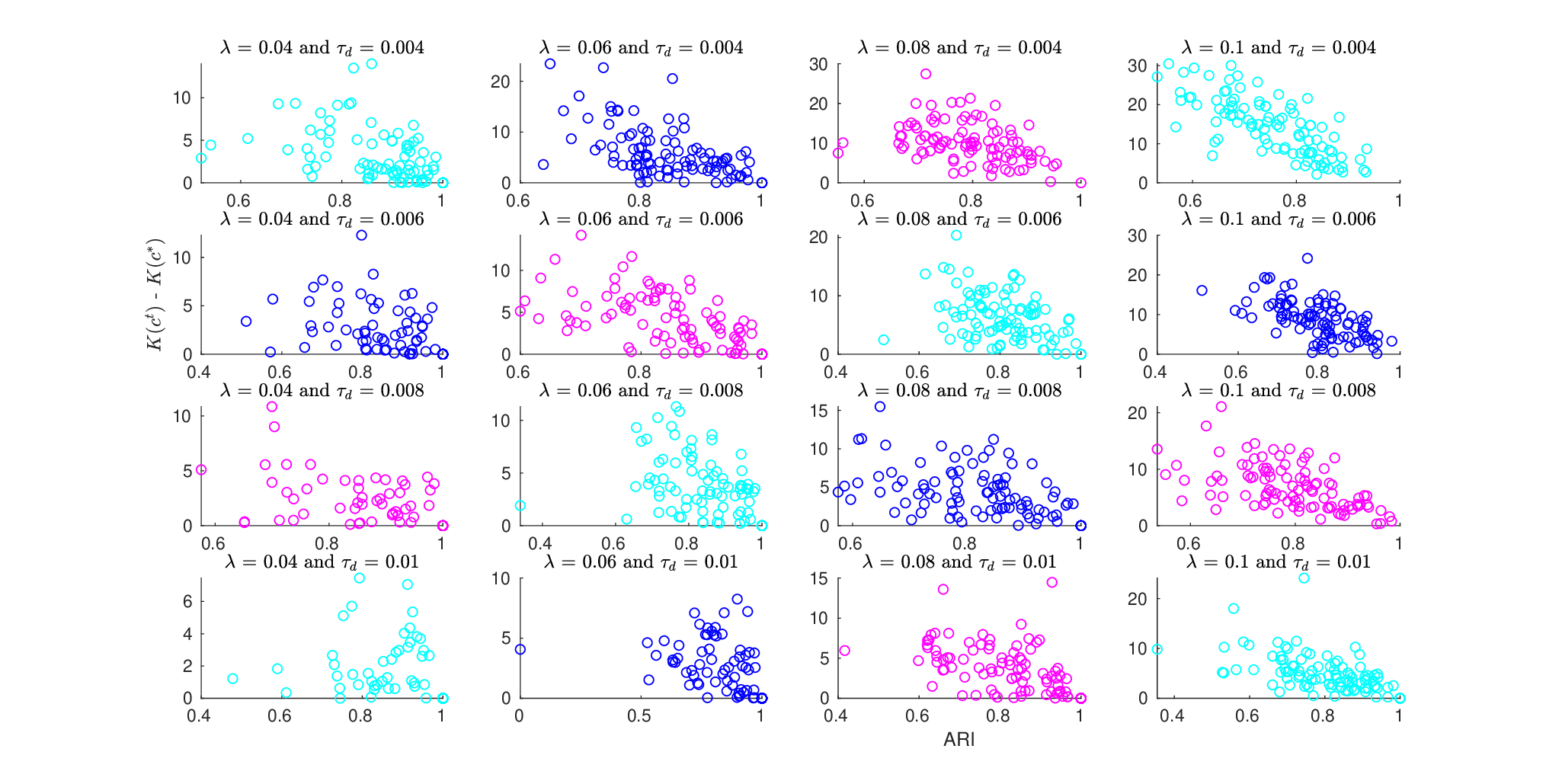}
\caption{The difference between $K(c^t)$ and $K(c^*)$ versus ARI for different values of birth rate and death rate.}
\label{fig:Analyse_ari}
\end{figure}
In this section, we evaluated the connection error caused by randomness. We display in Fig. \ref{fig:Analyse_ari} the scatter plots of ARI value vs $K(c^t)-K(c^*)$, where $c^*$ denotes the optimal configuration calculated by the "Tracklets Connection Algorithm" while $c^t$ is the true configuration. Each scatter plot displays the results of 100 simulations for a given combination of $\lambda$ and $\tau_d$. 

The difference between $K(c^t)$ and $K(c^*)$ is always positive or null, showing the optimization procedure works correctly to find the optimal solution. When $K(c^t)>K(c^*)$, it means that the configuration $c^*$ has a higher probability than the true realization $c^t$, which can be due to randomness. We can notice that ARI decreases as soon as $K(c^t)-K(c^*)$ increases. This error occurs when the realization is significantly different from the optimal configuration. Overall, ARI values are generally above 0.7.

Overall, we observe an increment of ARI when the difference decreases. The point clouds are diagonally shaped from top left to bottom right, showing a continuity that bigger the difference between $K(c^t)$ and $K(c^*)$, lower are the values of ARI. 

Therefore, to improve the performance of the connection model, it is needed to find some characteristics of different trajectories to distinguish the true realization. 
To investigate the connection errors, we display in Fig. \ref{fig:Analyse_ari} the scatter plots of ARI value vs. the difference between $K(c^t)$ and $K(c^*)$, where $c^*$ denotes the optimal configuration calculated by the "tracklets Connection Algorithm" while $c^t$ is the true configuration. Each scatter plot is associated with a given combination of $\lambda$ and $\tau_d$. We considered 100 replications for this experiment. On the left column, the connection is perfect in most case. Many points overlapped and  located at coordinates (1,0). Overall, we observe an increment of ARI when the difference decreases. This can be due to low values of ARI obtained when the ground truth does not correspond to the optimal configuration. 
\section{The boundary of $P(c)$}\label{appendix_2}
With the optimization algorithm (Eq. \ref{neme_sol}), we note $c_i$ the i\textsuperscript{th} optimal solution. Note that $N_c$ is the number of all possible configurations given $S$. For any $1\leq n\leq N_c$, we have
\begin{eqnarray*}
\sum_{i=1}^{n}Q(c_i)\leq\sum_{i=1}^{N_c}Q(c_i)\leq \sum_{i=1}^{n}Q(c_i) + (N_c-n)Q(c_{n})
\end{eqnarray*}
Using Eq. \ref{eq:LimEpsQ}, it gives
\begin{eqnarray*}
\frac{Q(c)}{\sum_{i=1}^{n}Q(c_i) + (\tilde{N}_c-n)Q(c_{n})}\leq P(c) \leq \frac{Q(c)}{\sum_{i=1}^{n}Q(c_i)},
\end{eqnarray*}
where $\tilde{N}_c$ is the number of possible configurations calculated through the number of inputs and outputs. We know that $\tilde{N}_c$ is bigger than $N_c$ because some configurations counted in $\tilde{N}_c$ are not compatible according to the time of inputs and outputs. Unfortunately we don't know the exact $N_c$.

\noindent Note $l(c,n)=\frac{Q(c)}{\sum_{i=1}^{n}Q(c_i) + (\tilde{N}_c-n)Q(c_{n})}$ and $u(c,n)=\frac{Q(c)}{\sum_{i=1}^{n}Q(c_i)}$. In order to guarantee the precision of the probability, we choose $n^*, 1\leq n^* \leq N_c$, big enough to satisfy, for $\alpha > 0$, 
\begin{eqnarray}\label{eq:left_bound}
\frac{(\tilde{N}_c-n^*)Q(c_{n^*})}{\sum_{i=1}^{n^*}Q(c_i)} < \alpha.
\end{eqnarray}
This gives us
\begin{eqnarray*}
\frac{u(c,n^*)}{l(c,n^*)}=\frac{\sum_{i=1}^{n^*}Q(c_i) + (\tilde{N}_c-n^*)Q(c_{n^*})}{\sum_{i=1}^{n^*}Q(c_i)}=1+\frac{(\tilde{N}_c-n^*)Q(c_{n^*})}{\sum_{i=1}^{n^*}Q(c_i)} < 1+\alpha
\end{eqnarray*}
which ensures that the upper and lower bounds are close from each other.
As a result, it gives
\begin{eqnarray*}
\sum_{i=1}^{n^*}P(c_i)>\sum_{i=1}^{n^*}l(c_i,n^*)>\frac{1}{1+\alpha}.
\end{eqnarray*}
In other words, this means that the set of configurations $c_i$ up to $n^*$ correspond to an highly likely event for $\frac{1}{1+\alpha}$ close to 1.
\section{Summary of notations useful for the evaluation of the likelihood}\label{sm:notations}

\begin{itemize}
\item $H$: the length of the cylinder,
\item $L$: the length of the circumference  of the cylinder,
\item $l$: the length of the part of the of circumference that is observed,
\item $l_u$: the length of the part of the of circumference that is not observed,
\item $l_e$: the difference between the length of the unobserved region and the observed region,

\item $B_c$: for a given configuration (reconstruction) of trajectories $c$,  the subset of trajectories born in the unobserved region and seen on the border $\{-l\}\times [0,H]$,
\item $D_c$: for a given configuration (reconstruction) of trajectories $c$,  the subset of trajectories seen on the border  $\{0\}\times [0,H]$ and died in the unobserved region,
\item $\Delta t$: time stepsize between two consecutive observations,

\item $N_l$: the number of particles born in the observed region and reaching the border  $\{0\}\times [0,H]$,
\item $p_x$: the probability of birth of a particle in a strip of width $x$, to the left side of the border  $\{-l\}\times [0,H]$,
\item $\hat{p}_x$: an estimator of $p_x$,
\item $S$: the observation set of all trajectories,
\item $S_o$: the set of tracklets having an output in $\{0\}\times [0,H]$,
\item $S_l^*$: the set of tracklets having an input in $\{-l\}\times [0,H]$ and an ouput in $\{0\}\times [0,H]$, that is crossing the observed region,
\item $S_r$: sample of points inside  a restricted region inside the observed region. This region should allow to decide if a particle died or is just moving outside the observed region.
\item $\tau_{\alpha}$: arrival rate at border $\{l\}\times [0,H]$ of particles born in the unobserved region $]-L, l[$,
\item $\hat{\tau}_{\alpha}$: estimator of the arrival rate  at border $\{l\}\times [0,H]$,
\item $\tau_d$: death rate of particles,
\item  $\hat{\tau}_d$: estimator of the death rate,
\item $T_S$: time duration of observation.

\end{itemize}